\documentclass[11pt, letter]{article}
\pdfoutput=1

\usepackage{jheppub}
\usepackage{natbib}

\usepackage[utf8]{inputenc}

\usepackage{mathtools}
\usepackage{dsfont}
\usepackage{mathdots}
\usepackage{amsmath}
\usepackage{amssymb}
\usepackage{amsthm}
\usepackage{mathrsfs}
\usepackage{slashed}
\usepackage{tabularx}
\usepackage{graphicx}
\usepackage{caption,subcaption}
\usepackage{float}

\usepackage{tikz}
\usetikzlibrary{arrows,shapes}
\usetikzlibrary{trees}
\usetikzlibrary{patterns}
\usetikzlibrary{matrix,arrows} 				%
\usetikzlibrary{positioning}				
\usetikzlibrary{calc,through}				
\usetikzlibrary{decorations.pathreplacing}  
\usepackage{pgffor}							

\usetikzlibrary{decorations.pathmorphing}	
\usetikzlibrary{decorations.markings}

\def\be{\begin{equation}}
\def\ee{\end{equation}}
\def\bea{\begin{eqnarray}}
\def\eea{\end{eqnarray}}
\def\ba{\begin{array}}
\def\ea{\end{array}}
\def\bd{\begin{displaymath}}
\def\ed{\end{displaymath}}

\def\>{\rangle} 
\def\<{\langle} 
\def\Dsl{D \hskip-.6em \raise1pt\hbox{$ / $ } }
\def\to{\rightarrow}

\preprint{LCTP-21-22}

\title{The Phase Diagram of BPS Black Holes in AdS$_5$}

\author{Nizar Ezroura,}
\author{Finn Larsen,}
\author{Zhihan Liu,}
\author{and Yangwenxiao Zeng}

\affiliation{Leinweber Center for Theoretical Physics,\\ 
Department of Physics, University of Michigan,\\
450 Church St, Ann Arbor, MI 48109, USA}
\emailAdd{nezroura@umich.edu}
\emailAdd{larsenf@umich.edu}
\emailAdd{billyliu@umich.edu}
\emailAdd{zengywx@umich.edu}

\abstract{Motivated by recent studies of supersymmetric black holes, we revisit the phase diagram of AdS black holes, whether BPS or not, with particular emphasis on the role of rotation. We develop BPS thermodynamics systematically and, in 
many explicit examples, we study its striking similarities with more familiar AdS black holes, as well as some differences. We highlight an important fugacity that preserves BPS saturation but is not captured by the supersymmetric index. 

\vspace{2cm} 
}


\begin{document}  

\maketitle

\newpage

\section{Introduction and Summary}
\label{sec:Introduction}

The thermodynamics of AdS black holes has been studied by numerous researchers \cite{ Hawking:1982dh, Chamblin:1999tk, Caldarelli:1999xj, Silva:2006xv, Benini:2015noa, Almheiri:2016fws, Hosseini:2016tor, Benini:2015eyy, Hosseini:2017mds, Benini:2016rke, Choi:2018hmj, Choi:2018vbz, Choi:2019miv, Cabo-Bizet:2018ehj, Larsen:2019oll, Copetti:2020dil, Larsen:2020lhg}.
However, in the last few years there has been a resurgence of interest in {\rm BPS} black holes, 
primarily focussed on the supersymmetric index of their dual CFT's \cite{Larsen:2019oll, Larsen:2020lhg, Cabo-Bizet:2018ehj, Choi:2018hmj, Benini:2018ywd}, and we are unaware of 
studies that interpolate between the very well known thermodynamics of generic AdS black holes
and their BPS limit. The basic raison d'\^{e}tre of this article is to address this gap in the literature, 
with emphasis on the complications due to angular momentum. It is a welcome bonus that 
this poorly explored corner of the black hole phase diagram turns out to 
illuminate questions that are central to ongoing microscopic studies of BPS black holes. 

BPS black holes are characterized, first of all, by their mass satisfying the BPS 
equality\footnote{In this article we set the AdS scale $\ell=1$ unless specifically noted. With this convention the black hole quantum numbers $M$, $Q_I$, $J_a$ are all integers for bosons while, for fermions, they are half-integral.}
\begin{equation}
\label{eqn:MBPS}
M = \sum_I Q_I + \sum_i J_i ~,
\end{equation}
relating their mass and charges. For AdS$_5$ the ranges $I=1, 2, 3$ and $i= a, b$ but the formula applies in other dimensions when ranges are adapted appropriately. In the context of a microscopic theory (or the Euclidean path integral) we can isolate such states by taking the extremal ($T\to 0$) limit 
of the partition function
\begin{equation}
\label{eqn:ZBPS}
Z = {\rm Tr} ~e^{-\beta (M - \Phi_I Q_I - \Omega_i J_i)} ~~\underset{T\to 0} \to~~
Z_{\rm BPS} = {\rm Tr} ~e^{\Phi'_I Q_I + \Omega'_i J_i} ~,
\end{equation}
where primes denote derivatives with respect to $T=\beta^{-1}$ and we have absorbed the supersymmetric Casimir Energy in the black hole mass. The limit $T\to 0$ is sufficient to project onto states that satisfy \eqref{eqn:MBPS}. That in turn is possible only for BPS states, due to unitarity, because the difference between the two sides of \eqref{eqn:MBPS} appears on the right hand side of the anticommutator between a supercharge and its conjugate. 

This setup is entirely standard, perhaps even boring, but it is also just the background for our story \cite{Kinney:2005ej, Choi:2018hmj}. Contemporary research on BPS black holes in AdS, whether in gravity or in CFT, nearly always imposes the additional condition on potentials that 
\begin{equation}
\label{eqn:varphip}
\varphi'= \sum_I \Phi'_I - \sum_i \Omega'_i \overset{?}{=} 0~,
\end{equation}
{\it vanishes}. 
There are excellent reasons to do so. The supersymmetric {\it index} is defined via the
BPS partition function \eqref{eqn:ZBPS}, with the grading $(-)^F$ with respect to fermion number inserted. The index is independent of temperature $T$, so the limit $T \to 0$ is optional, and the limit $T\to \infty$ may well be more convenient for computations. The reason that the index does not depend on $T$ is, as usual, that states violating the BPS bound are ``long",
in that the relevant supercharge acts nontrivially, and so contributions from bosons and fermions in the supermultiplet cancel. The brute force limit $T\to 0$ accomplishes this task as well, albeit with less elegance. 

However, the index is more powerful, it also ensures that the states that do satisfy the BPS bound cannot combine into long multiplets as continuous parameters, such as the coupling constant, vary. This only works if the fugacities assigned to BPS states are consistent with with the preserved supercharge, a condition that amounts precisely to $\varphi'=0$. Thus, if $\varphi'\neq 0$, the partition function generally depends on continuous parameters whether or not $(-)^F$ is inserted. Such loss of control discourages analysis in supersymmetric quantum field theory at weak coupling but, from the vantage point of the gravitational dual, there is no reason not to proceed. Quite the contrary, the AdS/CFT correspondence is routinely exploited to compute partition functions in strong coupling regimes that are inaccessible to other methods \cite{Larsen:2019oll, Larsen:2020lhg, Drukker:2010nc, Aharony:2003sx}. 

The processes whereby BPS states combine or disassociate into fewer or more BPS states preserve the linear mass relation 
$\eqref{eqn:MBPS}$ by rearranging the mass between $R$-charge $\sum_I Q_I$ and angular momentum $\sum_i J_i$. The index is, by construction, insensitive to such reorganizations but, for macroscopic black holes, it is fundamental that {\it all} conserved charges are specified. The dependence of black hole thermodynamics on $\varphi'$ amounts to a probe on how equilibrium forms along the chain of competing partitions of $(\sum_I Q_I )+ (\sum_i J_i)$ \cite{Larsen:2021wnu}.

To appreciate the central significance of this question, recall that the hope of describing BPS black holes in AdS in terms of weakly coupled CFT's has a history of disappointment that was only partially addressed by recent progress. It was long thought that the supersymmetric index was order ${\cal O}(1)$ in the rank $N$ of the underlying theory, unlike the exponential growth needed to account for the black hole entropy \cite{Kinney:2005ej, Berkooz:2006wc, Grant:2008sk, Chang:2013fba}. That would be unfortunate but there is no general argument against the cancellations brought about by insertion of $(-)^F$ being very efficient, effectively removing the leading asymptotics completely. Over the last few years evidence has emerged that such large cancellations can be prevented, after all, while preserving the protective abilities of the index, by introducing complex fugacities that compensate for the sensitivity to the statistics factor $(-)^F$ \cite{Cabo-Bizet:2018ehj, Choi:2018hmj, Benini:2018ywd, Choi:2018vbz, Choi:2019miv}, at least in part. Our study offers a complementary perspective on the dynamics along this direction in configuration space. 

As we noted already, early literature on the thermodynamics of AdS black holes did not highlight the BPS limit. The renewed interest in recent years was stimulated 
by the Hosseini-Hristov-Zaffaroni (HHZ) free energies \cite{Hosseini:2017mds}, and rightfully so. The AdS$_5$ version can be written as 
\begin{equation}
\label{eqn:HHZdef}
\ln Z = \frac{1}{2}N^2 \frac{\Delta_1 \Delta_2 \Delta_3}{\omega_a\omega_b}~,
\end{equation} 
where {\it complex} potentials $\Delta_I$, $\omega_{i}$ have real parts that can be identified with
$\Phi'_I$, $\Omega'_i$ but, in addition, they are endowed with complex parts. The HHZ free energy is at a central nexus because it can be derived, with some caveats, either from the gravitational action or from the supersymmetric index, and it may be related to the supersymmetric Casimir energy. As a point of comparison between such varied aspects of BPS black holes, it has a strict regime of applicability, required by the principles of supersymmetric localization. In particular, it is based on the index, the grading $(-)^{F}$ must be inserted, and it only applies when $\varphi'=0$. Therefore, by its very nature, it is not sensitive to the thermodynamic questions we study in this paper. We discuss the precise relation between the more conservative notion of gravitational free energy that we study and the index motivated HHZ expression in subsection \ref{subsubsec: HHZpota=b}.

The phase diagram for BPS black holes is surprisingly elaborate. It has been known for a few years that, with $\varphi'=0$, it is qualitatively similar to that of the AdS-Schwarzschild black hole \cite{Hawking:1982dh,Kubiznak:2016qmn}. There are small and large black hole branches joined by a cusp, and a Hawking-Page transition point to a thermal gas phase, all within a setting that does not break supersymmetry. Our work augments the BPS phase diagram and show that, in various ranges of the ``primed" BPS potentials, the small black holes may exhibit a maximal temperature, their free energy at high temperature may asymptote to a nonzero value, or the small black hole branch may disappear altogether. The large black holes, on the other hand, exhibit simple scaling laws at high temperature. These structures are reminiscent of the phase diagram for non-supersymmetric AdS black holes but here they appear in the context of BPS thermodynamics, justifying hope for a precision understanding of these interesting features starting from microscopic principles. 

The intricate phase diagram for BPS black holes, and its striking resemblance to analogous plots for generic AdS black holes, motivate a closer study of the entire phase diagram. In section \ref{sec:ads5thermodynamics} we explore the general thermodynamics of $\text{AdS}_5$ black holes in the grand canonical ensemble, including the effects of electric potential $\Phi$ and angular velocities $\Omega_{a,b}$. The critical values $\Phi^*=3$ and $\Omega^*_{a,b}=1$ correspond to BPS saturation. They give rise to phase boundaries where physical properties change qualitatively. We detail the behavior as one or more critical values is approached, while paying special attention to the orders of limits. 

All BPS black holes have the same potentials, they take their critical values, and vanishing free energy. In section \ref{sec:towardsbps} we introduce the BPS free energy $W$ and the ``primed" BPS potentials $\Phi'_I, \Omega'_i$ that it depends on. The thermodynamics of BPS black holes in AdS$_5$ is developed with emphasis on a ``benchmark" case, the simplest example. We also connect the gravitational thermodynamics to the dual CFT via the complex HHZ potential, establishing an exact match whether $(-)^F$ is inserted ($\varphi'=2\pi i$) or not ($\varphi'=0$). 

Finally, in section \ref{sec:BPSthermodynamics} we study the thermodynamics of BPS black holes for general BPS potentials. We determine the physical ranges of the BPS potentials and show that generic values of $\varphi'$ can modify the phase diagram qualitatively. Throughout this article our main example is the Kerr-Newman AdS$_5$ black hole and its BPS limit. By this we mean ``diagonal" charges $Q_I = Q$ and two angular momenta $J_a, J_b$ that may or may not be equal. In particular, we study $J_a \neq J_b$ for BPS black holes in subsection \ref{subsec:BPS0mu}.


\section{General Thermodynamics of $\text{AdS}_5$ Black Holes}
\label{sec:ads5thermodynamics}
In this section we study the thermodynamics of $\text{AdS}_5$ black holes for the case with equal charges ($Q_I = Q$) but general non-equal angular momenta $J_{a,b}$. 

Taking Schwarzschild-AdS as a benchmark, the phase diagram deforms smoothly as electric potential and angular velocities are added. Angular momentum proves more destabilizing than charge.

\subsection{Black Hole Thermodynamics}
\label{subsec:ads5thermo}

The Kerr-Newman AdS$_5$ family of black holes is characterized by four conserved charges $(M, Q, J_a, J_b)$. These physical charges are parametrized by $4$ variables $(m,q,a,b)$ through:
\begin{eqnarray}
\label{eqn: mass}
M & =&\frac{\pi}{4G_5}\frac{m(3- a^2g^2 - b^2g^2
- a^2 b^2g^4)+2qabg^2(2 - a^2g^2 - b^2g^2)}{(1-a^2g^2)^2 (1-b^2 g^2)^2}~,\\
\label{eqn: charge}
Q & = & \frac{\pi}{4G_5}\frac{q}{(1-a^2g^2)(1-b^2 g^2)}~,\\
\label{eqn: angular momentuma}J_a&=&\frac{\pi}{4G_5}\frac{2ma+qb(1+a^2g^2)}{(1-a^2g^2)^2 (1-b^2 g^2)}~,\\ 
\label{eqn: angular momentumb}J_b & = & \frac{\pi}{4G_5}\frac{2mb+qa(1+b^2g^2)}{(1-a^2g^2) (1-b^2 g^2)^2}~.
\end{eqnarray}
We will assume that all the conserved charges are nonnegative. Their scale   
is set by the dimensionless parameter 
\begin{equation*}
\label{eqn: dimlesspar}
\frac{\pi\ell^3_5}{4G_5} = \frac{1}{2}N^2~,
\end{equation*}
where $N$ refers to the dual $SU(N)$ gauge group of ${\cal N}=4$ SYM. The only other dimensionful parameter entering the thermodynamic formulae is the AdS$_5$ scale $\ell_5$ that is equivalent to the coupling of gauged supergravity $g=\ell^{-1}_5$. Henceforth we set $\ell_5=1$ (and so $g=1$), to avoid clutter, but the AdS$_5$ scale is easily restored. For example, $M, Q$ are inverse lengths, $a, b$ are lengths, and $q, m$ are lengths squared.

The event horizon is at the coordinate location $r_+$, a combination of parameters that is ubiquitous in thermodynamic formulae for AdS black holes. 
It is determined as the largest root of the horizon equation
\begin{eqnarray}
\label{eqn: ehorizon}
\label{Delta_r} &&\Delta_r(r)=\frac{(r^2+a^2)(r^2+b^2)(1+r^2)+q^2+2abq}{r^2}-2m=0~.
\end{eqnarray}
With $r_+$ given implicitly through this equation, the black hole temperature is
\begin{eqnarray}
\label{eqn:T}
&&T=\frac{r_+^4[1+(2r_+^2+a^2+b^2)]-(ab+q)^2}{2\pi r_+[(r_+^2+a^2)(r_+^2+b^2)+abq]}~,
\end{eqnarray}
and the electric potential and angular velocities are
\begin{eqnarray}
\label{eqn:Phi}
\Phi &=& \frac{3qr_+^2}{(r_+^2+a^2)(r_+^2+b^2)+abq}~,
\\
\label{eqn: Oma}
\Omega_a &=&\frac{a(r_+^2+b^2)(1+r_+^2)+bq}{(r_+^2+a^2)(r_+^2+b^2)+abq}~,\\ \Omega_b &=&\frac{b(r_+^2+a^2)(1+r_+^2)+aq}{(r_+^2+a^2)(r_+^2+b^2)+abq}~.
\label{eqn:Omb}
\end{eqnarray}

The linchpin for thermodynamics is, of course, the black hole entropy computed from the area law: 
\begin{eqnarray}
\label{eqn:S}
&&S=2\pi\cdot  \frac{1}{2}N^2\cdot \frac{(r_+^2+a^2)(r_+^2+b^2)+abq}{(1-a^2)(1-b^2)r_+}~,
\end{eqnarray}

The conserved charges 
(\ref{eqn: mass}-\ref{eqn: angular momentumb}) and the thermodynamic potentials (\ref{eqn:T}-\ref{eqn:S}), with the subsidiary condition \eqref{Delta_r} that determines $r_+$ implicitly, completely determine the thermodynamics of Kerr-Newman AdS$_5$ black holes. The nonlinearities evident in these parametric equations contain a great deal of physics. 

The black hole mass $M$ is bounded from below by the BPS mass 
\begin{equation*}
\label{eqn: bpsmass}
M_{\rm BPS} = \Phi^* Q + \Omega^*_a J_a  + \Omega^*_b J_b~,
\end{equation*}
where $\Phi_*=3$ and $\Omega^*_a = \Omega^*_b = 1$. In this section we consider generic variables, a detailed study of BPS thermodynamics follows later. However, we will find that the {\it critical} values of the potentials, denoted by stars, play a central role also away from the BPS limit. In particular, we consider only $\Omega_{a,b}\leq \Omega^*_{a,b}=1$, because the ``overspinning" black holes with larger values are classically unstable due to superradiance.

\subsection{The Grand Canonical Ensemble}
\label{subsec:GCE}
In this subsection we study the phase diagram in the grand canonical ensemble, i.e. as a function of temperature $T$, electric potential $\Phi$, and angular velocities $\Omega_{a,b}$. These potentials correspond to periodicities of the (Euclidean) time, the gauge function and two azimuthal angles, respectively. Therefore, the grand canonical ensemble is the natural description in the path integral formalism where asymptotic boundary conditions are specified as geometrical data. 

The Gibbs free energy is given by
\begin{eqnarray}
\label{gibbsfe}
  G &=& M - TS - \Phi Q - \Omega_a J_a - \Omega_b J_b~.
\end{eqnarray}
The extensive variables $M$, $Q$, $J_{a,b}$, $S$ are presented in
(\ref{eqn: mass}-\ref{eqn: angular momentumb},
\ref{eqn:S}).
After trading $m$ for $r_+$ through the horizon equation \eqref{eqn: ehorizon} they become functions of the parameters $(r_+, q, a, b)$. We can explicitly eliminate $q$ in favor of the potential $\Phi$ 
\begin{eqnarray}
\label{eqn:qtoPhi}
    q&=&\frac{\Phi (r_+^2+a^2)(r_+^2+b^2)}{3r_+^2-ab\Phi}~,
\end{eqnarray}
by inverting \eqref{eqn:Phi}. However, the parameters $(r_+,a,b)$ must remain as implicit functions of the potentials $T$, $\Phi$ and $\Omega_{a,b}$ because it is impractical to solve (\ref{eqn:T}, \ref{eqn: Oma}-\ref{eqn:Omb}). Roughly, but not precisely, the horizon location $r_+$ is a proxy for the temperature $T$, and the rotation parameters $a, b$ represent rotation velocities $\Omega_{a,b}$. 

After the elimination of the parameter $q$, the angular velocities become 
\begin{eqnarray}
    \label{eqn:ExpansionOa}
    1-\Omega_a & = & \frac{1-a}{r_+^2+a^2}
    \left[r_+^2-r_\ast^2+b(1+a)(1 - \frac{1}{3} \Phi)\right]~,
    \\
    \label{eqn:ExpansionOb}
    1-\Omega_b &=& \frac{1-b}{r_+^2+b^2}\left[r_+^2-r_\ast^2+a(1+b)(1 - \frac{1}{3}\Phi)\right]~.
\end{eqnarray}   
These formulae are completely general, albeit presented in a manner that anticipate a special role for the BPS horizon position $r^2_\ast=\sqrt{a+b+ab}$ and the critical values of the potentials $\Omega^*_a=\Omega^*_b=1$, $\Phi^*=3$. In a similar spirit, we can present the general temperature as
\begin{equation}
    \begin{split}
        T=&\frac{r_+(r_+^2+a+b-ab)}{2 \pi \left(r_+^2-a b\right)}
        \left( \frac{1 - \Omega_a}{1-a} + \frac{1 - \Omega_b}{1-b}\right) \\
        &-\frac{r_+}{6\pi}\frac{\left(3+\Phi-3a-3b\right)r_+^2+(a^2+b^2+a^2b+ab^2)\Phi-ab(\Phi-3)}{(r_+^2-ab)(3r_+^2-ab\Phi)}\left(\Phi-3\right)~.
    \end{split}
    \label{eqn:ExpandT}
\end{equation}
In this form it is manifest that the temperature vanishes when the potentials take their critical values. We are not able to explicitly invert the three equations (\ref{eqn:ExpansionOa}-\ref{eqn:ExpandT}) any further but, taken together, they relate the physical potentials $(T, \Omega_a, \Omega_b)$ to the parameters $(r_+, a, b)$ for a given $\Phi$.

After elimination of the parameter $q$ in favor of the electric potential $\Phi$, Gibbs' free energy $G$ becomes:
\begin{equation}
    \begin{split}
        G=&-\frac{N^2}{4}\frac{(r_+^2+a^2)(r_+^2+b^2)(r_+^2+a+b-ab)(r_+^2-r_\ast^2)}{(1-a^2)(1-b^2)(r_+^2-ab)^2}\\
        &-\frac{N^2\left(r_+^2+a^2\right) \left(r_+^2+b^2\right)\left(\Phi-3\right)}{12 (1-a^2)(1-b^2) \left(r_+^2-a b\right)^2 \left(3 r_+^2-a b \Phi\right)^2}\left[(r^2_+-a b)^2 
        ( 9r^2_+ +  3r^2_+\Phi -2 ab \Phi^2)\right. \\
        &~~~~~~~~~~ \left. - 3(a+b)^2 r^2_+ ( 3r^2_+ + r^2_+\Phi - 2ab\Phi) \right]~.
    \end{split}
  \label{eqn:generalG}
\end{equation}
Again, we have pulled out factors that make it manifest that the expression vanishes when the potentials take their critical values. 

In the following subsections we study the phase diagram for various ranges or special values of $\Phi$, with any $\Omega$, and for two exceptional values of $\Omega$, for any $\Phi$. When combined, these special cases map out the entire phase diagram. 

\subsubsection{Angular Velocity $\Omega$ with Electric Potential $\Phi=0$}
\label{subsubsec:phi0}
\begin{figure}
    \centering
    \includegraphics[width=0.8\textwidth]{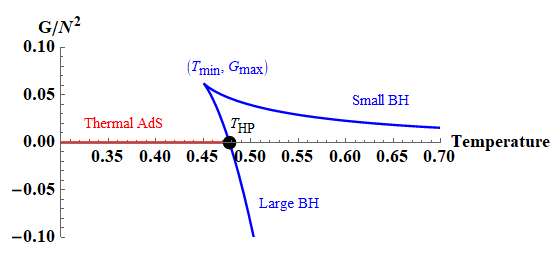}
    \caption{AdS-Schwarzschild phase diagram ($\Phi=0,~\Omega_{a,b}=0$). The cusp is at the minimal temperature $T_{\text{min}}=\frac{\sqrt{2}}{\pi}$, and the maximal free energy $G_{\text{max}}=\frac{N^2}{16}$.
    The large black hole branch meets the thermal AdS phase at the Hawking-Page temperature $T_{\text{HP}}=\frac{3}{2\pi}$.} 
    \label{fig:phase00}    
\end{figure}

If, in addition to taking the electric potential $\Phi=0$, we also consider vanishing angular velocities $\Omega_{a,b}=0$, we are left with the Schwarzschild-AdS black hole. In this case the phase diagram, reproduced in Figure \ref{fig:phase00}, is extremely well-known \cite{Witten:1998qj}. Some of its features are:    
\begin{itemize}
    \item
    There is a strictly positive lower bound on the black hole temperature $T\geq T_{\text{min}}$.
    \item
    For each temperature $T\geq T_{\text{min}}$ there are two branches. Black holes on the upper one are ``small" in units of the AdS$_5$ radius. They are qualitatively similar to asymptotically flat black holes. 
    The ``large" black holes on the lower branch are influenced significantly by the AdS$_5$ background. 
    \item
    A thermal AdS phase has the same boundary conditions at infinity as the black holes but vanishing free energy. It dominates when the temperature is below the Hawking-Page temperature $T_{\text{HP}}$ \cite{Hawking:1982dh}, where the large black hole also has zero free energy. The large black holes are thermodynamically preferred only when they have temperature $T\geq T_{\text{HP}}$. 
\end{itemize}
The Hawking-Page transition is particularly interesting because it can also be interpreted as the confinement/deconfinement transition in QCD-like theories \cite{tHooft:1977nqb, Polyakov:1978vu, Susskind:1979up,Aharony:2003sx} living on the boundary via the AdS/CFT correspondence\cite{Witten:1998zw, Witten:1998qj, Maldacena:1999}.

As the angular velocities $\Omega_{a,b}$ are turned on and increased, with the electric potential kept at $\Phi=0$, the qualitative features of the AdS-Schwarzschild phase diagram are preserved. Naturally, there are quantitative changes: 
\begin{itemize}
\item
For any given $T$, on either branch, the angular velocities lower the free energy. This follows from the first law that gives  $\partial_{\Omega_{a,b}} G = - J_{a,b}<0$.
\item
The transition temperature $T_{\text{HP}}$ decreases as either angular velocity increases, as it must because the large black hole branch is lowered.
\item
The minimal black hole temperature $T_{\text{min}}$ also decreases as either angular velocity increases. The maximal free energy, attained at the cusp, increases. Thus the cusp travels ``towards North-West" when angular velocity increases.
\item
In the limit where the angular velocities $\Omega_{a,b}\to 1$
the height of the cusp diverges $G_{\text{max}} \to \infty$ but
its temperature approaches a finite (nonzero) limit $T_{\text{min}}\to \frac{1}{2\pi}$ . In the strict limit $\Omega_a=\Omega_b=1$ the ``large" branch disappears altogether. This limit is detailed in subsection \ref{subsubsec:Omega=1}. 
\end{itemize}
The existence of an absolute minimum for the temperature for any angular velocities and the increase in maximal free energy both suggest that, everything else being equal, rotation tends to {\it destabilize} the black hole. 

When $\Phi = 0$, the two angular velocities are ``decoupled" in that 
\begin{equation}
    \label{eqn:Oaphi=0}
    \Omega_a=\frac{a(1+r_+^2)}{r_+^2+a^2}~,
\end{equation}
is a function only of $r_+$ and $a$, but not $b$.
The formula for $\Omega_b$ is analogous. 

The free energy \eqref{eqn:generalG} also simplifies greatly when $\Phi=0$:
\begin{eqnarray}
    G=-\frac{N^2}{4}\frac{\left(r_+^2-1\right) \left(r_+^2+a^2\right) \left(r_+^2+b^2\right)}{r_+^2\left(1-a^2\right) \left(1-b^2\right)}~.
    \label{eqn:Gphi=0}
\end{eqnarray}
The Hawking-Page transition, where $G=0$, corresponds to the horizon parameter $r_+=1$. With this value we can invert 
\eqref{eqn:Oaphi=0} (and its analogue for $\Omega_b$) and find $a$ ($b$) as a function of $\Omega_a$ ($\Omega_b$). These expressions, along with $r_+=1$, give a simple formula for the Hawking-Page transition temperature 
\begin{equation}
    T  ~\underset{\Phi=0}{=}~
    \frac{r_+^4 \left(2 r_+^2+a^2+b^2+1\right)-a^2 b^2}{2 \pi  r_+ \left(r_+^2+a^2\right) \left(r_+^2+b^2\right)}
 ~  \underset{\rm HP}{=}~
\frac{1+\sqrt{1-\Omega_a^2}+\sqrt{1-\Omega_b^2}}{2\pi}~.
\label{eqn:THP}
\end{equation}
It is a decreasing function of the two angular velocities independently, as expected. It interpolates between the AdS-Schwarzschild value $T_{\rm HP} = \frac{3}{2\pi}$ when there is no rotation, and approaches the finite value $T_{\rm HP}=\frac{1}{2\pi}$ towards $\Omega_a=\Omega_b=1$. where G diverges. We discuss the subtle limiting case in subsection \ref{subsubsec:Omega=1}. 

\begin{figure}[htb]
        \centering
        \includegraphics[width=0.8\linewidth]{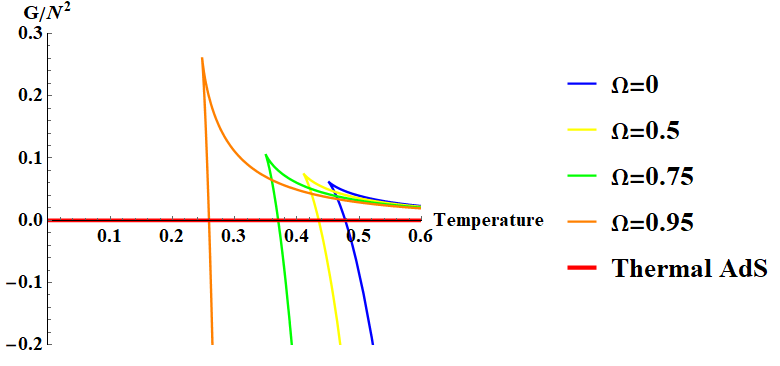}
        \caption{Gibbs' free energy $G$ as function of the temperature $T$. The electric potential vanishes $\Phi=0$ and the angular velocities increase from right to left as $\Omega_a=\Omega_b=0,~0.5,~0.75,~0.95$. The cusp moves ``North-West" as angular velocities increase, staying above the minimum temperature $T_{\rm min}= \frac{1}{2\pi} \approx 0.16$ even in the limit $\Omega_a=\Omega_b\to 1$.}
        \label{fig:phase1}
\end{figure}

The phase diagram for vanishing potential $\Phi=0$ and a sample of angular velocities is presented in Figure 
\ref{fig:phase1}.

There has been suggestions that the ``thermal gas in AdS" 
that competes with the black hole phase may not exist, except for AdS-Schwarzschild. The free energy is evaluated by a gravitational path integral with asymptotic boundary conditions that are satisfied by a black hole, but also compatible with a bulk spacetime that has no black hole. 
The latter has $G=0$ (up to a Casimir term) and, whatever its precise physical nature, it must be taken into account. It is on this basis that we will consider black holes with $G>0$ unstable, also for spacetimes with angular velocity and/or electric potential. 


\subsubsection{Subcritical Electric Potential: $0<\Phi<3$}
\label{subsubsec:phil3}

\begin{figure}[htb]
    \centering 
    \includegraphics[width=0.8\linewidth]{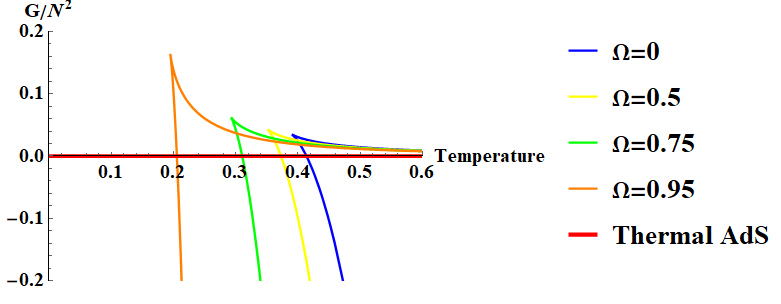}
    \caption{Gibbs' free energy $G$ as function of the temperature $T$. The electric potential $\Phi=1.5$ is in the range $0<\Phi<3$ and the angular velocities increase from right to left as $\Omega_a=\Omega_b=0,~0.5,~0.75,~0.95$. For a given $\Omega_{a,b}$, the free energy and the temperature are both smaller than for $\Phi=0$ (compare with Figure \ref{fig:phase1}.)}
    \label{fig:phase2}
\end{figure}

Starting from vanishing electric potential $\Phi$, but arbitrary angular velocities (below their critical values $\Omega_a^*=\Omega_b^*=1$), we now consider increasing the electric potential toward its critical value $\Phi^*=3$.
The phase diagram remains qualitatively similar to the AdS-Schwarzschild case as the electric potential increases, but quantitative features are modified. Some of these changes are similar to an increase in angular velocity at fixed electric potential:
\begin{itemize}
    \item 
The free energy at a given temperature decreases, on both the small and the large black hole branches. This follows from the thermodynamic relation $\partial_\Phi G= - Q$. 
\item
The transition temperature $T_{\text{HP}}$ decreases, as it must because the entire large black hole branch is lowered. 
The minimal temperature, attained at the cusp where the two branches meet, also decreases. 
\end{itemize}
However, despite these similarities, increasing electric potential differs significantly from increasing angular velocity:
\begin{itemize}
\item
The free energy at the cusp $G_{\rm max}$ {\it decreases} with increasing electric potential. 
\item 
The transition temperature $T_{\text{HP}}$ (and so the minimal temperature $T_{\rm min}<T_{\text{HP}}$) decreases with {\it no minimum nonzero value} as the electric potential approaches the critical value $\Phi=\Phi^*$.
\end{itemize}
Both of these features indicate that increased potential $\Phi$ {\it stabilizes} the black hole. The phase diagram with electric potential in the range $0<\Phi<3$ is presented in Figure \ref{fig:phase2}.
    
\subsubsection{Critical Electric Potential $\Phi=\Phi^*=3$}
\label{subsubsec:phi3}
When $\Phi$ increases to the critical value $\Phi^*=3$, the phase diagram changes {\it qualitatively}. 

The expressions for potentials 
     \begin{eqnarray}
     \label{eqn:Tphi=3}
     &&T|_{\Phi=3}=\frac{r_+(2r_+^2+a^2+b^2)\left(r_+^2+a+b-ab\right)}{2\pi(r_+^2+a^2)(r_+^2+b^2)(r_+^2-ab)}(r_+^2-r_*^2)~,\\
     \label{eqn:Oaphi=3}
     &&1 - \Omega_a|_{\Phi=3}=\frac{1-a}{r^2_++a^2}(r^2_+-r_*^2)~,\\
     \label{eqn:Obphi=3}
     && 1 - \Omega_b|_{\Phi=3}=\frac{1-b}{r^2_++b^2}(r^2_+-r_*^2)~,
    \end{eqnarray}
and for Gibbs free energy
 \begin{eqnarray}
     &&G|_{\Phi=3}=-\frac{N^2 (r_+^2+a^2)(r_+^2+b^2)\left(r_+^2+a+b-ab\right)}{4(1-a^2)(1-b^2)(r_+^2-ab)^2}(r_+^2-r_*^2)~,
     \label{eqn:Gphi=3}
    \end{eqnarray}
simplify somewhat. As in earlier formulae,  $r_*=\sqrt{a+b+ab}$ is the horizon location for BPS black holes with parameters $(a,b)$.
    
The equations show that, when $\Phi=\Phi^*=3$, the requirement of positive temperature $T\geq0$ is equivalent to $r_+\geq r_*$.
Therefore, taking $\Phi=\Phi^*=3$ automatically prevents overspinning (it keeps $\Omega_{a,b}\leq1$) and it also ensures non-positive free energy $G\leq0$. Because of these inequalities, the phase diagram simplifies greatly when $\Phi=\Phi^*$. There is {\it no small black hole branch}, no matter what the temperature and angular velocities are. In particular, there is no ``cusp" where two branches meet. Additionally, since black hole states have non-positive free energy, the thermal AdS gas with $G=0$ is never preferred. 

When any of the bounds noted in the previous paragraph are saturated we have $r^2_+=r^2_*$ and so they all become equalities. Moreover, in this case the black hole is BPS. In other words, the following are equivalent: 
\begin{itemize}
    \item $\Phi=\Phi^*$ and {\it any} of $T=0$, $\Omega_{a}=1$, $\Omega_{b}=1$, $G=0$. 
    \item $\Phi=\Phi^*$ and {\it all} of $T=0$, $\Omega_{a}=1$, $\Omega_{b}=1$, $G=0$. 
    \item The black hole is {\it supersymmetric}, i.e. the mass is exactly the BPS mass: 
    $M = M_{\rm BPS}= \Phi^* Q + \Omega^*_a J_a + \Omega^*_b J_b$.
\end{itemize}

The formulae (\ref{eqn:Tphi=3}-\ref{eqn:Obphi=3}) suggest a specific {\it approach} to the BPS limit: simply fix the parameters $a, b$ and tune black hole parameters so that  $r^2_+-r^2_*=\epsilon\to 0$. This corresponds to physical potentials approaching the BPS limit {\it linearly} as
\begin{eqnarray}
\label{eqn:Tlim}
    &&T|_{\Phi=3} 
    =\frac{(2+a+b)\sqrt{a+b+ab}}{\pi(1+a)(1+b)(a+b)}\epsilon~,\\
    \label{eqn:Omalim}
    &&1 - \Omega_a|_{\Phi=3}=\frac{1-a}{(1+a)(a+b)}\epsilon~,\\
    \label{eqn:Omblim}
    && 1 - \Omega_b|_{\Phi=3}=\frac{1-b}{(1+b)(a+b)}\epsilon~,
\end{eqnarray}
with Gibbs free energy \eqref{eqn:generalG} also approaching zero linearly
\begin{eqnarray}
\label{eqn:Gphi3zero}
     &&G|_{\Phi=3}=-
     \frac{1}{2} N^2 \frac{a+b}{(1-a)(1-b)}\epsilon~.
\end{eqnarray}
When implementing the BPS limit in this way it is manifest that $G, T, 1-\Omega_{a,b}$ all reach $0$ simultaneously, with specified relative rate. In section \ref{sec:towardsbps} we will identify this approach 
with the BPS limit itself. 

\begin{figure}[htb]
\centering 
    \includegraphics[width=\linewidth]{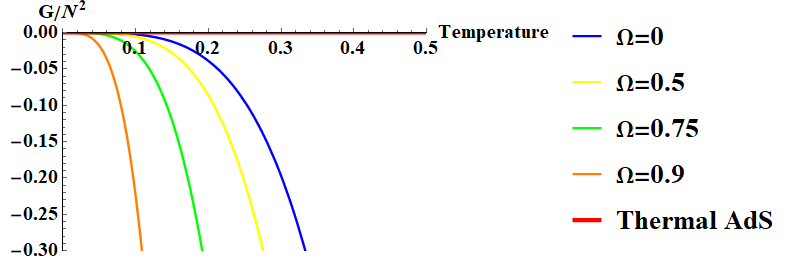}
    \caption{The free energy $G$ as function of temperature $T$ for $\Phi=\Phi^*=3$. The angular velocity $\Omega_a=\Omega_b=0,~0.5,~0.75,~0.95$ from right to left. The curves approach the origin $G=T=0$ {\it without reaching it}. The special case $\Omega_a=\Omega_b=1$ is not a curve, it is the single point $G=T=0$.}
    \label{fig:phase3}
\end{figure}

The somewhat formal statements on the BPS limit in the previous paragraphs fail to reflect all physical aspects.  Therefore, we now consider the physical realization of thermodynamics already discussed for $\Phi<3$: start with high temperature and then cool the system all the way to $T=0$, while keeping the physical angular velocities $\Omega_{a,b}$ fixed. Figure \ref{fig:phase3} shows the resulting free energy $G$ as function of temperature $T$, for various values of $\Omega_a=\Omega_b$. 

To implement the physical thermodynamics underlying Figure \ref{fig:phase3} we would lower the parameter $r_+$, a proxy for temperature, and simultaneously lower the parameters $a,b$, in order that the left hand sides of (\ref{eqn:Oaphi=3}-\ref{eqn:Obphi=3}) remain constant. 
The approach to the BPS limit, described 
in (\ref{eqn:Tlim}-\ref{eqn:Omblim}), is 
different because it keeps the parameters $a, b$ fixed. 
The distinction is dramatic because Figure \ref{fig:phase3} depicts a smooth approach to $T=G=0$ for various $\Omega_{a,b}$, in apparent contradiction with the statement that vanishing temperature $T=0$ is possible only when $\Omega_a=\Omega_b=1$ identically. 

To resolve the tension, it is sufficient to consider the regime $r^2_+ \gg a^2, b^2, ab$ where formulae simplify so that: 
\begin{align}
T & = \frac{r^4_+  - (a+b)^2}{\pi r^3_+}~,
\cr
\frac{1 - \Omega_a}{1-a}  & =  \frac{1 - \Omega_b}{1-b}  =  1 - \frac{ a + b}{r^2_+}~,
\cr
 G &= - \frac{1}{2} N^2 \frac{r^4_+ -(a+b)^2}{(1-a^2)(1-b^2)}~.
 \label{eqn:simpleGab}
\end{align}
We further take $a, b \ll 1$ so $r^2_+ \sim a+b$ (which easily satisfies $r^2_+ \gg a^2, b^2, ab$) in a manner where the rotational velocities can take any value $\Omega_{a,b} = \frac{a+b}{r^2_+}$, as long as $\Omega_a = \Omega_b$. Then 
\begin{equation}
\label{eqn:lowTapx}
T = \frac{1}{\pi} r_+ ( 1 - \Omega^2)~,
\end{equation}
and the free energy becomes: 
\begin{equation}
\label{eqn:Gsimpleq}
G = -  \frac{1}{2} N^2 \frac{\pi^4 }{(1 - \Omega^2)^3}T^4 ~.
\end{equation}
These formulae are not the most general, but they exhibit important features clearly. 

The BPS limit (\ref{eqn:Tlim}-\ref{eqn:Omblim}) lowers $r_+$ so $r^2_+-r^2_*=\epsilon\to 0$. In this case (\ref{eqn:lowTapx}-\ref{eqn:Gsimpleq}) are 
realized by $G$, $T$, $1 - \Omega^2$ all approaching zero at the same rate, while $r_+$ is near the constant $r_*$.

The low temperature limit with the angular velocity $\Omega$ fixed is also described by the simplified free energy \eqref{eqn:Gsimpleq} but, due to \eqref{eqn:lowTapx}, the strict limit $T\to 0$ requires
$r^2_+\to 0$ (with $a, b\to 0$ such that $\Omega_{a,b} = \frac{a+b}{r^2_+}$ is fixed). The limiting geometry with $r_+=0$ is singular, so it is not actually a solution. 
It is a ``small" black hole, not just in the AdS/CFT vernacular, where the term refers to black hole size that is much smaller than the AdS-scale $r_+\ll \ell_5$, but in the sense that unknown higher derivative curvature corrections dominate the classical ``solution".
Therefore, the BPS limit is well-controlled only if it is taken as specified in (\ref{eqn:Tlim}-\ref{eqn:Omblim}).

The free energy \eqref{eqn:Gsimpleq} is reminiscent of the {\it high} temperature regime $r_+\gg 1$.
In this limit the rotational velocities $\Omega_a = a$, $\Omega_b = b$ , the temperature \eqref{eqn:Tphi=3} simplifies to 
$T=\frac{r_+}{\pi}$ and the free energy \eqref{eqn:simpleGab} becomes 
\begin{equation}
\label{eqn:Gphi3LargeT}
    G= - \frac{1}{2}N^2\frac{\pi^4}{(1-\Omega_a^2)(1-\Omega_b^2)} T^4 ~.
\end{equation}
This formula shares with the low temperature expression \eqref{eqn:Gsimpleq} the power $T^4$ that is characteristic of CFT's in four dimensions. However, factors of $1-\Omega^2$ differ. 
The high temperature regime is dual to the conformal fluid \cite{Bhattacharyya:2007vjd, Rangamani:2009xk}.

\subsubsection{Super-critical Electric Potential $\Phi>3$}
\label{subsubsec:phig3}

\begin{figure}[htb]
  \includegraphics[width=\linewidth]{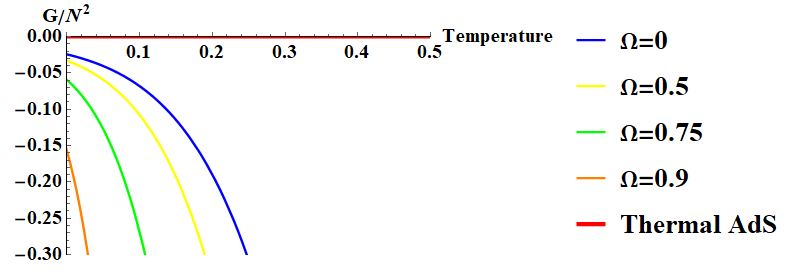}
  \caption{The free energy $G$ when $\Phi=3.5$. The angular velocities $\Omega_a=\Omega_b=0,~0.5,~0.75,~0.9$ increase from right to left.}
  \label{fig:phase4}
\end{figure}

When $\Phi$ increases above the critical value $\Phi_*=3$ all black holes acquire strictly negative free energy. Therefore, they are thermodynamically preferred to thermal AdS$_5$. The free energy decreases monotonically as $\Phi$ increases further. Moreover, it decreases further as the angular velocities increase with fixed $\Phi$ and $T$. All these trends follow straightforwardly from the thermodynamic relations $\partial_\Phi G = - Q$ and $\partial_{\Omega_{a,b}} G = - J_{a,b}$. They are illustrated in Figure \ref{fig:phase4}.

 To be more specific, we consider the case $\Omega_a=\Omega_b \equiv \Omega$, corresponding to $a=b$. Then we can invert 
 the angular velocity \eqref{eqn:ExpansionOa} and solve for $r_+^2$
\begin{equation}
\label{eqn:rpaOm}
    r_+^2=\frac{a(3(1-a\Omega)+\Phi(1-a^2))}{3(\Omega-a)}~,
\end{equation}
and the temperature \eqref{eqn:ExpandT} becomes
\begin{equation}
\label{eqn:TaPhiOmrp}
    T=\frac{a(3+\Phi-6\Omega^2)-(\Phi-3)\Omega}{6\pi a(1-a\Omega)}r_+~.
\end{equation}
The nontrivial solution for extremality ($T=0$)  
\begin{equation}
\label{eqn:extremala}
    a=\frac{(\Phi-3)\Omega}{3+\Phi-6\Omega^2}~,
\end{equation}
yields the free energy \eqref{eqn:generalG} in terms of $\Phi$ and $\Omega_a=\Omega_b=\Omega$ at extremality: 
\begin{equation}
    G_{\text{ext}}=-\frac{N^2\left(\frac{1}{3}\Phi-1\right)^2}{16(1-\Omega^2)^2}\left[3\left(\frac{1}{3}\Phi+1\right)^2-4\Omega^2\left(1+\frac{2}{3}\Phi\right)\right].
    \label{eqn:GT=0}
\end{equation}
This analytical formula gives the strictly negative free energy at $T=0$ for any $\Phi>3$. In the nonrotating limit $\Omega\to 0$ (further discussed in the following subsection) the formula is regular, so non-rotating black holes in AdS$_5$ are stable when the electric potential is large. This result applies also in the flat space limit where it is significant for studies of the weak gravity conjecture (WGC) \cite{Harlow:2021trr, Harlow:2022gzl}. 

We are particularly interested in the BPS limit which requires $\Omega=1$. When $\Phi>3$ the free energy is singular for this value: the physical range of the angular velocity $\Omega$ is $[0,1)$, so $\Omega$ can be arbitrarily close to the critical value $\Omega^*=1$, but it cannot reach it. However, the square bracket in \eqref{eqn:GT=0} vanishes when both $\Omega=1$ and $\Phi=3$. Therefore, in the limit $\Phi\to 3$, the singularity at $\Omega=1$ is suppressed by three powers of $\frac{1}{3}\Phi-1$. 
This shows that the free energy vanishes $G\to 0$ when $\Omega\to1$ and $\Phi\to3$ at identical ``speed". 


The reason that the angular velocities $\Omega_{a,b}$ can be arbitrarily close to their critical values $\Omega^*_{a,b}=1$, but cannot reach it, 
is that the conserved charges $M, Q, J_{a,b}$ must be finite: they can be arbitrarily large, but they cannot diverge. This in turn requires $a,b\in[0,1)$, neither of these parameters can be exactly equal to one. 

To see this, first consider the temperature \eqref{eqn:ExpandT} near $a=1$ and/or $b=1$:
\begin{eqnarray}
    \label{eqn:Tabnear1}
   T  \underset{a=1-\epsilon_a,b=1-\epsilon_b}{=} \frac{r_+\Big(r_+^2-(\frac{2}{3}\Phi+ 1)- \frac{1}{18}(\Phi^2-9)\Big)}{\pi\left(r_+^2-\frac{\Phi}{3}\right)}+\mathcal{O}(\epsilon_a, \epsilon_b)~.
\end{eqnarray}
Nonnegativity of the temperature in this regime shows
$r_+^2\geq (\frac{2}{3}\Phi+ 1)>0$. 
Therefore, since $\epsilon_{a,b}$ can be small but not strictly zero, the angular velocities \eqref{eqn:ExpansionOa} and \eqref{eqn:ExpansionOb}
\begin{align}
\label{eqn:Omanear1}
    &1-\Omega_a\underset{a=1-\epsilon_a,b=1-\epsilon_b}{=}\frac{\left(r_+^2-(\frac{2}{3}\Phi+1)\right)}{(1+r_+^2)}\epsilon_a+\mathcal{O}(\epsilon^2)~,\\
\label{eqn:Ombnear1}
    &1-\Omega_b\underset{a=1-\epsilon_a,b=1-\epsilon_b}{=}\frac{\left(r_+^2-(\frac{2}{3}\Phi+1)\right)}{(1+r_+^2)}\epsilon_b+\mathcal{O}(\epsilon^2)~.
\end{align}
can be arbitrarily close, but not equal to, the critical values $\Omega^*_{a,b}=1$. In other words, when $\Phi>3$ either of the 
limits $\Omega_{a}\to 1$ or $\Omega_{b}\to 1$ takes {\it all} the quantum numbers $M, Q, J_{a, b}\to\infty$


\subsubsection{Nonrotating Black Holes: $\Omega=0$}
\label{subsubsec:nonrotbhs}

In this subsection we turn off rotation by setting $a=b=0$ and focus on the effect of the electric potential $\Phi$. This case was well developed early on \cite{Chamblin:1999tk,Caldarelli:1999xj} so it serves as an important benchmark for the effects of rotation. 

\begin{figure}
    \centering
    \includegraphics[width=0.8\textwidth]{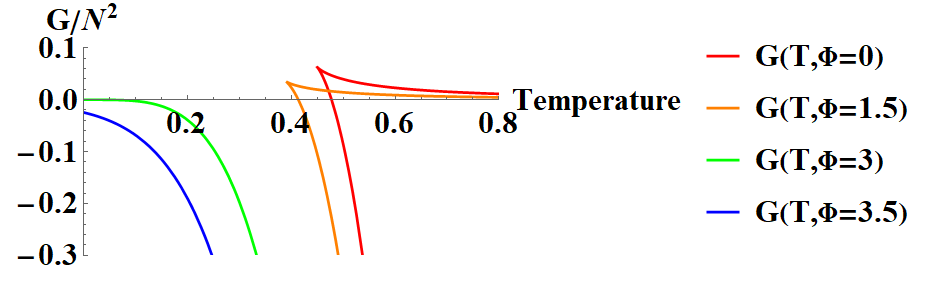}
    \caption{The free energy for non-rotating black holes $\Omega=0$. The electric potential $\Phi$ increases from right to left. There is a qualitative transition from AdS-Schwarzschild type at $\Phi < \Phi_*=3$ to the high potential regime $\Phi\geq\Phi_*=3$.}
    \label{fig:phase_6_0}
\end{figure}

For zero angular velocity $\Omega_a=\Omega_b=0$, the 
temperature \eqref{eqn:ExpandT} and the free energy \eqref{eqn:generalG} reduce to 
\begin{align}
\label{eqn:TPhiZeroRot}
    T=&\frac{2 g^2r_+^2+1-\left(\frac{1}{3}\Phi\right)^2}{2\pi  r_+}~,\\
\label{eqn:GPhiZeroRot}
    G=&-\frac{N^2}{4}  r_+^2 \left(g^2r_+^2+\left(\frac{1}{3}\Phi\right)^2-1\right)~.
\end{align}
For given $(T, \Phi)$, the first equation yields $2$, $1$, or $0$ solutions for $r_+$; and then the second equation gives the applicable values of the free energy $G(T, \Phi)$. It is presented in Figure~\ref{fig:phase_6_0}. 

The AdS-Schwarzschild black hole, reviewed in the beginning of 
subsection \ref{subsubsec:phi0} and plotted in Figure \ref{fig:phase00}, is the curve to the right. Increasing electric potential $\Phi$ lowers both the minimal temperature $T_{\rm min}$ and the maximal free energy $G_{\rm max}$. These effects make the black hole more stable. 

There is a qualitative change as the potential increases from the regime $\Phi<3$ to $\Phi\geq 3$. The lower range of potential $\Phi<3$ is ``Schwarzchild-type", it has two black hole branches that are joined at a cusp where the temperature reaches its minimum $T_{\rm min}$ and the free energy its maximum $G_{\rm max}$. The higher range of potential $\Phi\geq 3$ has just a "large" black hole branch. Therefore, this range corresponds to particularly stable black holes, as identified by the weak gravity conjecture. 

The earlier parts of this subsection considered the effect of rotation on various ranges of $\Phi$. In particular, each curve in Figure~\ref{fig:phase_6_0} is the same as first curve shown in Figures \ref{fig:phase1}-\ref{fig:phase4}. The results of those earlier subsections showed that angular velocities strictly below maximal $0\leq\Omega_{a,b}<1$ do not change the phase diagrams qualitatively. For $\Phi<3$, angular velocity lowers the temperature $T_{\rm min}$ but the free energy $G_{\rm max}$ increases. For $\Phi > 3$ rotational velocity lowers the free energy of the large black hole branch, the only one there is. The case where the electric potential is exactly critical $\Phi=\Phi^*=3$ is subtle, as discussed in subsection \ref{subsubsec:phi3}.

Black hole thermodynamics is much simpler in the absence of rotation so more formulae can be made explicit. In this case, we can invert \eqref{eqn:TPhiZeroRot} and solve for
the coordinate location of the horizon $r_+$:
\begin{equation}
    r_+=\frac{\sqrt{2g^2 (\Phi^2-9)+(3 \pi T)^2}+3 \pi  T}{6 g^2}~.
    \label{eqn: horizonZeroRot}
\end{equation}
We picked the {\it positive} root. 
Then \eqref{eqn:GPhiZeroRot} gives the free energy: 
\begin{equation}
    G=-\frac{N^2}{4}\frac{\left(\sqrt{2g^2 \left(\Phi^2-9\right)+(3\pi T)^2 }+3 \pi  T\right)^2 \left(\pi  T \left(\sqrt{2g^2   \left(\Phi^2-9\right)+(3\pi T)^2}+3 \pi  T\right)+ g^2 \left(\Phi^2-9\right)\right)}{216g^6}~.
    \label{eqn:GTZeroRot}
\end{equation}
We restored the AdS$_5$ length scale $g=\ell_5^{-1}$ in order to facilitate comparison to the flat space limit $g\to0$. When $\Phi<\Phi_\ast=3$, the argument of the square roots may become negative so, as we have elaborated on extensively, there is a minimal temperature. For $\Phi>\Phi_\ast=3$, the case discussed in the previous subsection, there is a lower bound $r_+ > \frac{1}{6}\ell_5\sqrt{2(\Phi^2-9)}$ and the free energy is strictly negative in the limit. 

It is instructive to take $\Phi=\Phi_\ast=3$ from the outset and then take $T\to 0$ with the AdS scale $\ell_5\to\infty$ such that the product $T\ell_5=\Delta$ is fixed. In this limit the coordinate horizon 
\begin{equation}
    r_+=\pi T\ell^2_5=\pi \Delta \ell_5~,
\end{equation}
diverges. In this limit $\ell_5\to\infty$ the underlying black hole geometry is not "small", the position of the coordinate horizon $r_+$ approaches a finite limit, and here this corresponds to finite spatial extent. Indeed, the underlying "large" black hole geometry reduces to the asymptotically flat Reissner-Nordstr\"{o}m black hole. 

The low temperature regime with no rotation also makes contact with a large body of recent literature that studies universal features shared with the SYK model (some entry points to the literature \cite{Maldacena:2016upp,Sachdev:2019bjn}). In this context general thermodynamic arguments show that the approach to extremality is quadratic in temperature
 \cite{Preskill:1991tb,Almheiri:2016fws} and it is interesting that the coefficient of the quadratic temperature dependence can be computed directly in the extremal ($T=0$) geometry \cite{Larsen:2018iou,Hong:2019tsx}. 
Here the low temperature of the free energy \eqref{eqn:GTZeroRot} is {\it linear} in the temperature
\begin{equation}
    G= -\frac{N^2 \left(\Phi^2-9\right)^2}{432 g^2}-\frac{N^2 (\Phi^2-9)^{\frac{3}{2}}}{54\sqrt{2} g^3}\pi T-\frac{ N^2 \left(\Phi^2-9\right)}{24 g^4}\left(\pi T\right)^2+ \ldots
    \label{eqn:GTZnattr}
\end{equation}
for a given $\Phi>\Phi_* = 3 $, 
There is an apparent tension, but it is resolved because our study is in the grand canonical ensemble (fixed potentials), while the quadratic behavior pertains to the canonical ensemble (fixed charges).


\subsubsection{Maximal Rotational Velocity: $\Omega=1$} 
\label{subsubsec:Omega=1}

It is instructive to consider black holes with maximal horizon velocities $\Omega_{a}=\Omega_{b}=1$ for any temperature $T$ and electric potential $\Phi$. This example shows that generally such spins are destabilizing,  but the BPS case is an exception. 

The general expressions for $\Omega_{a,b}$  (\ref{eqn:ExpansionOa}-\ref{eqn:ExpansionOb}) give the horizon position  
\begin{equation}
\label{eqn:rpome1}
r_+^2=r_\ast^2 - b(1+a)(1 - \frac{1}{3}\Phi)~.
\end{equation}
It also gives the analogous expression with $a\leftrightarrow b$ so, for consistency, the rotational parameters must be identical $a=b$.
The general formula for temperature \eqref{eqn:T} becomes  
\begin{eqnarray}
\label{eqn:TPhia}
T=& \frac{1}{2\pi} (1-\frac{1}{3}\Phi) \sqrt{( 1+\frac{1}{3}\Phi)a^{-1}+ \frac{1}{3} \Phi}~.
\end{eqnarray}
Since the temperature is nonnegative $T\geq0$, we must demand $\Phi\leq3$. This agrees with the analysis in subsection \ref{subsubsec:phig3}: the rotational velocities can not reach their maximal values $\Omega_{a,b}=1$ when $\Phi>3$. 

In most of the examples we consider, the horizon location $r_+$ is {\it larger} than the corresponding BPS radius $r_*$, at the same values of $a, b$. One might wonder if this is a physical condition. For example, it could be that black holes that are excited above the BPS limit must be bigger because they have more entropy. On the other hand, the parameter $r_+$ is not a physical variable, nor are $a,b$, so this comparison cannot be taken too seriously. Indeed, according to \eqref{eqn:rpome1}, black holes with $\Omega=1$ give a counterexample because they have $r_+<r_*$. 

The temperature $T$ given in \eqref{eqn:TPhia} is a monotonically decreasing function of the parameter $a$. Since black holes with finite charges must have $a<1$, the temperature is bounded from below by 
the value at $a=1$:
\begin{equation}
 T_{\rm min}= \frac{1}{2\pi}
(1-\frac{1}{3}\Phi) \sqrt{1+\frac{2}{3}\Phi}~.
    \label{eqn:Tmin}
\end{equation}
The inequality $a<1$ is strict, $a=1$ is not possible, so the lower bound $T \geq T_{\rm min}$ can be saturated only when $\Phi=3$. In fact, when $\Phi=3$, the minimal temperature $T_{\rm min}=0$ is not just a bound, 
\eqref{eqn:TPhia} shows it is the only possible temperature. That the temperature must vanish in this case is precisely as expected, because when $\Phi=3$, $\Omega_{a,b}=1$ all potentials are critical and so the black hole is BPS $M=M_{\rm BPS}$. Here we focus on the interesting discontinuity between $\Phi\to 3^-$ and $\Phi=3$.

\begin{figure}
    \centering
    \includegraphics[width=0.8\textwidth]{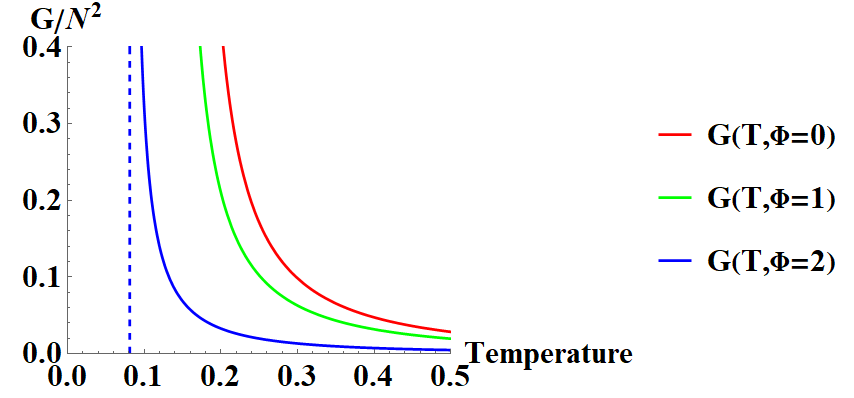}
    \caption{Free energy as function of temperature when $\Omega_{a,b}=1$ 
    and $\Phi \neq 3$. There are {\it only ``small" black holes}. They all have $G>0$ and negative specific heat ($G$ decreases with temperature). The curves represent $\Phi=0,1,2$ (from right to left). The vertical asymptote indicating the minimal temperature is shown only for $\Phi=2$, to reduce clutter. As $\Phi$ increases to $\Phi\to 3_-$ the minimal temperature decreases to $T_{\rm min}=0$. The BPS limit where $\Phi=3$ exactly is not a curve, it is just the origin where $T=0$ and $G=0$.}
    \label{fig:phase7}
\end{figure}

For maximal rotational velocities $\Omega_{a,b}=1$ and any generic potential $0\leq \Phi<3$, all temperatures $T>T_{\rm min}$ are possible.  Gibbs' free energy \eqref{gibbsfe} becomes:
\begin{equation}
\begin{split}
    G &=\frac{1}{2} N^2\frac{  a (3-\Phi) (\Phi+3)^2}{54 (1-a)}
    \cr
 &=\frac{1}{4}N^2\frac{\left(1 -\frac{1}{9} \Phi^2\right)^3}{(2\pi T)^2 - ( 1 + \frac{2}{3}\Phi) ( 1 - \frac{1}{3}\Phi)^2}~.
\end{split}
\label{eqn:GOm1Phil3}
\end{equation}
In the second expression the rotational parameter $a$ was eliminated by taking advantage of \eqref{eqn:TPhia}. 
This free energy, plotted in Figure \ref{fig:phase7}, is positive definite, so the corresponding black holes are always ``small". The minimal temperature $T_{\rm min}$ given in \eqref{eqn:Tmin} is such that the denominator is positive for all temperatures $T> T_{\rm min}$ and vanishes in the limit $T\to T_{\rm min}$. Therefore, the free energy diverges in this limit. 

The challenges faced by the approach to the BPS limit $T=0$, $\Phi=3$ are stark when $\Omega_{a,b}=1$ is set from the outset: 
\begin{itemize}
    \item 
There is a minimal temperature $T_{\rm min}$. However, it decreases monotonically as a function of the potential $\Phi$, from $T_{\rm min}=\frac{1}{2\pi}$ when $\Phi=0$ to $T_{\rm min}\to 0$ as $\Phi\to 3_-$. 
\item
The free energy diverges if, for a given potential, the temperature $T$ is lowered to $T_{\rm min}$. However, the ``height" of the divergence is set by $(1 - \frac{1}{3}\Phi)^3$. 
\item
By increasing $\Phi\to 3_-$ and lowering $T$ simultaneously, the free energy can vanish in the limit, after all. However, the free energy necessarily approaches zero from {\it above} $G\to 0_+$.
\end{itemize}
The ``safest" approach to the BPS limit are from the large black hole branch $G\to 0_-$. It requires taking electric potential $\Phi=\Phi_\ast=3$ first, or at least approaching it from above $\Phi\to 3_+$, and only then tuning other parameters.

\section{Thermodynamics of Supersymmetric Black Holes}
\label{sec:towardsbps}

In this section we study the thermodynamics of black holes that are strictly supersymmetric. We consider general aspects, as well as their realization by black holes in AdS$_5$. As an instructive example, we develop new details on a  ``benchmark" case that was discussed previously. The following section \ref{sec:BPSthermodynamics} studies more general cases. 

\subsection{General Structure of BPS Thermodynamics}
\label{subsec:genBPSthermo}

Suppose we have a general free energy\footnote{In this subsection we refer to a single electric potential $\Phi$ and a single angular velocity $\Omega$, to keep terminology simple. However, one or both kinds of potentials can have multiple components, corresponding to several independent charges and/or angular momenta.} 
\begin{equation}
\label{eqn:genfreenergy}
G = G(T, \Phi, \Omega) = M- TS - \Phi Q - \Omega J~,
\end{equation}
and want to study the states that satisfy the BPS mass relation
\begin{equation}
\label{eqn:ads4MBPS}
M = \Phi^* Q + \Omega^* J~,
\end{equation}
where $\Phi^*$ and $\Omega^*$ are given numbers that depend on the system we consider. The general free energy \eqref{eqn:genfreenergy} satisfies the 
first law
\begin{equation*}
\label{eqn:firstlaw}
dG = - S dT - Q d\Phi - Jd\Omega~,
\end{equation*}
so it yields the mass 
\begin{equation*}
\label{eqn:massfirstlaw}
M  = G + TS + \Phi Q  + \Omega J = (1 - T\partial_T  - \Phi\partial_\Phi - \Omega\partial_\Omega) G~.
\end{equation*}
The BPS mass relation \eqref{eqn:ads4MBPS} is satisfied if and only if
\begin{equation*}
\label{eqn:Ghomcond}
\Big[1 - T\partial_T G - (\Phi-\Phi^*)\partial_\Phi - (\Omega-\Omega^*)\partial_\Omega\Phi\Big] G = 0 ~.
\end{equation*}
This is the condition that the free energy is {\it homogeneous of degree $1$} in the variables $T, \Phi-\Phi^*,\Omega-\Omega^*$. 

Considering any $G$ that is homogeneous of degree $1$, we define the BPS free energy
\begin{equation}
W(\Phi',\Omega')= \frac{G}{T} = - S - \Phi'  Q - \Omega' J~.
\label{eqn:FreeEnergy}
\end{equation}
The second form used the BPS mass relation (\ref{eqn:ads4MBPS})
and introduced primed potentials
\begin{align}
\label{eqn:primPO}
\Phi' & =  \frac{\Phi-\Phi^*}{T}~,  \cr
\Omega'& =  \frac{\Omega-\Omega^*}{T}~.
\end{align}
The function 
W inherits a homogeneity property from $G$ and, because we have divided by $T$, it is in fact precisely homogeneous in the variables $T, \Phi-\Phi^*,\Omega-\Omega^*$, it has ``weight" zero. Therefore, it is possible to
interpret it as a function of the ratios $\Phi', \Omega'$, as we have indicated in our notation for $W$.
In the BPS limit we necessarily have temperature $T\to 0$ so these ``primed" variables can be identified as derivatives with respect to the temperature, evaluated at $T=0$. An alternative but equivalent interpretation is that the BPS surface is a projective manifold defined by ratios of thermodynamic variables.  

The BPS free energy $W$ satisfies the first law
\begin{equation}
\label{eqn:wdiff}
dW  = - Q  d \Phi'  - J d \Omega' ~, 
\end{equation}
so
\begin{align}
\label{eqn:wderi}
    Q = -\frac{\partial W}{\partial \Phi'} ~, &~~~~~~~ J = -\frac{\partial W}{\partial \Omega'}~.
\end{align}
It yields the entropy 
\begin{equation} 
\label{eqn:BPSentro}
 S =   ( - 1 + \Phi'\partial_{\Phi'} + \Omega'\partial_{\Omega'}) W~.
\end{equation} 
However, the BPS mass has been ``integrated out" of the BPS thermodynamics, it is not encoded in $W$ and only follows from $Q$ and $J$ via the BPS mass relation \eqref{eqn:ads4MBPS}. 

In statistical mechanics, the BPS free energy $W$ arises when introducing a BPS partition function as a trace over all BPS states
\begin{equation}
\label{eqn:ZBPSdef}
Z_{\rm BPS} = {\rm Tr}_{\rm BPS} \left[ e^{\Phi' Q + \Omega' J}\right]~,
\end{equation}
with no explicit reference to mass. Then the BPS free energy $W = - \ln Z_{\rm BPS}$.  

\subsection{Parameters for BPS Black Holes in AdS$_5$}
\label{subsec:BPS BH labels}

In this paper we primarily study $\text{AdS}_5$ black holes with a single charge and two independent angular momenta. We want to make the general considerations in the preceding subsection explicit in this specific case. 

The BPS condition is 
\begin{equation}
\label{eqn: bpscond}
    M = 3Q + J_a  + J_b~. 
\end{equation}
Comparing with the generic formula \eqref{eqn:ads4MBPS} we have 
\begin{equation}
\label{eqn:Phistar}
\Phi^*=3~,
\end{equation}
and, for $a=b$, $\Omega^*=2$. In the general case where $a\neq b$ there are two independent rotational velocities $\Omega_{a,b}$ with BPS values 
\begin{equation}
\label{eqn:Omegastar}
\Omega^*_a=\Omega^*_b=1~.
\end{equation}

The general parametric expressions for the black hole quantum numbers $(M, Q, J_{a, b})$ in (\ref{eqn: mass} - \ref{eqn: angular momentumb}) give the mass excess above the BPS bound 
\begin{equation}
\label{eqn:BPSm}
   M - 3Q - (J_a+J_b)  = \frac{1}{2}N^2  \frac{3+(a+b)-ab}{(1-a)(1+a)^2(1-b)(1+b)^2} \Big( m-q(1+a+b)\Big)~.
\end{equation} 
The prefactor on the right hand side is manifestly positive for all $a, b\in[0,1[$, so the BPS formula \eqref{eqn: bpscond} yields the condition on the parameters $(m,q,a,b)$:
\begin{equation}
\label{eqn: bpscond2}
    m = q(1+a+b)~.
\end{equation}
We can therefore consider $m$ a dependent variable and parametrize the BPS black holes by $(q, a, b)$.

Even when the parameter $m$ satisfies \eqref{eqn: bpscond2}, the equation \eqref{eqn: ehorizon} for the coordinate location of the event horizon is a cubic equation in $r^2$. Instead of proceeding by brute force, it is useful to recall that BPS black holes are extremal and so we expect a double root. With this clue, and the BPS 
condition \eqref{eqn: bpscond2}, it is manageable to find the solution 
\begin{equation}
\label{eqn:rstardef}
    r_* ^2 = a+b +ab ~,
\end{equation}
for a specific value of the parameter $q$:
\begin{equation}
\label{eqn:qstardef}
q^* = (1+a)(1+b)(a+b)~.
\end{equation}
Whichever way this solution for particular values of $m, q$ was arrived at, we can, for any $m, q$, recast the cubic equation \eqref{eqn: ehorizon} satisfied by the horizon location $r^2$ as
\begin{equation}
\label{eqn:rr*qq*}
    (r^2 - r^2_*)^2 r^2 + \Big( (q-q^*) - (r^2-r^2_*)(1 + a + b)\Big)^2 
    + 2r^2 (m - q( 1 + a + b)) = 0~.
\end{equation}
When the BPS condition \eqref{eqn: bpscond2} is satisfied this expression is the sum of two squares so, for BPS black holes, it is manifest that a real solution exists {\it only} for $q=q^*$ and for this value of $q$ there is indeed a double root when $r^2=r^2_*$. We will refer to the starred variables as the BPS values. 

When $m$ satisfies the BPS condition \eqref{eqn: bpscond2} and $q$ takes its BPS value \eqref{eqn:qstardef}, the parametric formulae (\ref{eqn: mass} - \ref{eqn: angular momentumb}) for the physical variables $(M, Q, J_{a,b})$ 
simplify to:
\begin{align}
 \label{eqn:BPSM}
    M^* &= \frac{1}{2}N^2 \frac{(a+b)(3 - a^2 + ab - b^2 - ab(a+b))}{(1-a)^2 (1-b)^2}~,\\
     \label{eqn:BPSQ}
    Q^* &= \frac{1}{2}N^2  \frac{a+b}{(1-a)(1-b)} ~,
   \\ 
     \label{eqn:BPSJa}
    J_a ^* &= \frac{1}{2}N^2  \frac{(a+b)(2a+b+ab)}{(1-a)^2 (1-b)} ~,\\
     \label{eqn:BPSJb}
    J_b ^* &= \frac{1}{2}N^2  \frac{(a+b)(a+2b+ab)}{(1-a) (1-b)^2}~. 
\end{align}
The potential $\Phi$ and the angular velocities $\Omega_{a,b}$ given in (\ref{eqn:Phi} - \ref{eqn:Omb}) reduce to the BPS values (\ref{eqn:Phistar}-\ref{eqn:Omegastar}), as they should.
The entropy $S$ \eqref{eqn:S} simplifies to:
\begin{equation}
    \label{eqn:BPSS}
    S^* = 2\pi \frac{1}{2}N^2 \frac{(a+b)\sqrt{a+b+ab}}{(1-a)(1-b)}~.
\end{equation}

As we have stressed, once we impose the BPS mass formula \eqref{eqn: bpscond}, black holes exist only when the parameters $q, a, b$ satisfy the relation \eqref{eqn:qstardef}. Thus the BPS mass cannot be reached for all black hole variables $(Q, J_a, J_b)$. Rather, according to \eqref{eqn:BPSQ}, \eqref{eqn:BPSJa} and \eqref{eqn:BPSJb}, the conserved charges must satisfy the constraint 
\begin{equation} 
\label{eqn:chargeconstraint}
\left(Q^{*3} +\frac{1}{2}N^2J_a^* J_b^* \right) - \left(3Q^*+\frac{1}{2}N^2 \right) \left(3Q^{*2} -\frac{1}{2}N^2 (J^* _a +J^* _b) \right)= 0~.
\end{equation} 
It is surprising that there is this condition on the charges. 

\subsection{The BPS Free Energy}
\label{subsec:approBPS}

Usually, a good starting point for the study of thermodynamics is the free energy $G$, an extensive function of the intensive thermodynamic potentials $T, \Phi, \Omega_{a,b}$. However, for BPS black holes the free energy vanishes identically $G=0$ and the ``variables" it would ordinarily be a ``function" of are fixed at $T=0, \Phi^*=3$, and $\Omega^*_{a,b}=1$. 

The general structure of BPS thermodynamics discussed in subsection \ref{subsec:genBPSthermo} addresses this obstacle. 
A free energy $G$ that is homogeneous of degree one in $T$, $\Phi-\Phi^*$, $\Omega_{a,b}-\Omega_{a,b}^*$ depends on variables that nominally depart from their BPS values, but the homogeneity property shows that the mass is {\it exactly} the BPS mass. Therefore, a free energy of this form describes thermodynamics intrinsic to the BPS surface. 

\subsubsection{Derivation of the BPS Free Energy}
\label{subsubsec:BPSWderiv}

In the AdS$_5$ example we focus on, the explicit free energy $G$ \eqref{eqn:generalG} of a general AdS$_5$ black hole depends on the electric potential $\Phi$, as it should, but it is also a function of the parameters $r_+, a, b$, rather than $T$, $\Omega_{a,b}$. Conveniently, the expression for $G$ given
in \eqref{eqn:generalG} vanishes when, in addition to $\Phi=3$,
the auxiliary horizon position $r^2_+$ take the 
BPS value $r^2_*$ given in \eqref{eqn:rstardef}. At linear order in 
both $r^2_+-r^2_*$ and $\Phi-3$ we have 
\begin{equation*} 
\label{eqn:GlinordrPhi}
G = - \frac{N^2}{2} \frac{a+b }{(1-a)(1-b)}(r^2_+-r^2_*)
 - \frac{N^2}{6}\frac{(a+b)(1-a-b-ab)}{(1-a)(1-b)}  (\Phi-3)~.
\end{equation*}
The combination $r^2_+-r^2_*$ has a geometrical interpretation in terms of horizon position, but it is not a traditional thermodynamic potential. It is related to the angular velocities through the two equations (\ref{eqn:ExpansionOa}-\ref{eqn:ExpansionOb}) which are straightforward
to expand at linear order in $\Phi-\Phi^*$, $\Omega_{a,b}-\Omega_{a,b}^*$. The result 
\begin{equation}
\label{eqn:r+rstar}
r^2_+-r^2_* = -\frac{(a+b)(1+a)}{2(1-a)} (\Omega_a - 1) - \frac{(a+b)(1+b)}{2(1-b)} (\Omega_b - 1)+\frac{a+b+2ab}{6}(\Phi-3)~,
\end{equation}
leads to the free energy
\begin{equation}
\label{eqn:Glinfct}
    G = -\frac{N^2}{12}\frac{(a+b)(2-a-b)}{(1-a)(1-b)} (\Phi -3) + \frac{N^2}{4}\frac{(a+b)^2(1+a)}{(1-a)^2 (1-b)} (\Omega_a - 1) +  \frac{N^2}{4}\frac{(a+b)^2(1+b)}{(1-a) (1-b)^2} (\Omega_b - 1)~.
\end{equation}
Moreover, the condition that the two equations 
(\ref{eqn:ExpansionOa}-\ref{eqn:ExpansionOb}) give identical values for $r^2_+-r^2_*$ 
yields the {\it constraint} on the potentials
\begin{equation}
    \frac{(a+b)(1+a)}{1-a}(\Omega_a-1) -\frac{(a+b)(1+b)}{1-b}(\Omega_b-1) + \frac{1}{3}(a-b)(\Phi-3) = 0 ~. 
    \label{eqn:constraint1}
\end{equation}
The formula for temperature \eqref{eqn:ExpandT}, also expanded
at linear order in $\Phi-\Phi^*$, $\Omega_{a,b}-\Omega_{a,b}^*$, 
imposes a second constraint on the potentials 
\begin{equation}
    T=-\frac{\sqrt{a+b+ab}}{\pi(1-a)}(\Omega_a -1) -\frac{\sqrt{a+b+ab}}{\pi(1-b)}(\Omega_b -1) -\frac{\sqrt{a+b+ab}}{3\pi(a+b)}(\Phi-3) ~.
    \label{eqn:constraint2}
\end{equation}
We can interpret the two constraints 
(\ref{eqn:constraint1}-\ref{eqn:constraint2}) together, as definitions of the auxiliary parameters $a, b$. Since both constraints are invariant under simultaneous linear scaling of $T, \Phi-\Phi^*$, $\Omega_{a,b}-\Omega_{a,b}^*$, 
the $a, b$ defined implicitly this way are homogeneous functions of these potentials. 

The free energy \eqref{eqn:Glinfct} transforms linearly under simultaneous linear scaling of $T, \Phi-\Phi^*$, $\Omega_{a,b}-\Omega_{a,b}^*$ so, as a function of these variables, it is homogeneous of degree one. According to the discussion in subsection \ref{subsec:genBPSthermo}, it follows that it describes BPS black holes. To reach this conclusion it is important that the parameters $a, b$ are invariant under such rescalings. 

To avoid possible confusion, we reiterate the reasoning. The free energy \eqref{eqn:Glinfct} and the constraints (\ref{eqn:constraint1}-\ref{eqn:constraint2}) were all derived by expanding to linear order in $\Phi-\Phi^*$, $\Omega_{a,b}-\Omega_{a,b}^*$. That could leave the impression that they are approximations that are valid near the BPS limit but in fact they describe the BPS black holes themselves, these equations contain no information beyond BPS. Once a free energy is presented as a homogeneous function, the additional step of taking the limit $T\to 0$ is possible, but not required, doing so corresponds to a choice of coordinates in projective geometry. 

To understand how this is possible, consider the mass excess over the BPS mass \eqref{eqn:BPSm}, rewritten using (\ref{eqn:rr*qq*}, \ref{eqn:r+rstar}): 
\begin{equation}
\label{eqn:MMstarTP}
    M-M_* = \frac{N^2}{2}\frac{(3+a+b-ab)(a+b)^2}{4(1-a)(1-b)(1+3a+3b+a^2+3ab+b^2)} \Big( (2\pi T)^2 + \varphi^2 \Big) ~,
\end{equation}
where 
\begin{equation}
\label{eqn: varphidef}
    \varphi = (\Phi-\Phi^*) -(\Omega_a - \Omega_a ^*) -(\Omega_b - \Omega_b ^*) = \Phi - \Omega_a -\Omega_b -  1~. 
\end{equation}
This mass excess $M-M_*$ is {\it quadratic} in the small variables and so the
black hole is BPS because the equality $M=M_*$ holds at linear order.\footnote{The quadratic correction in \eqref{eqn:MMstarTP} is not important in this article but it is interesting in its own right \cite{Preskill:1991tb,Almheiri:2016fws,Moitra:2018jqs,Nayak:2018qej,Larsen:2019oll,Iliesiu:2020qvm,Heydeman:2020hhw, Larsen:2020lhg}. It is described in effective quantum field theory by the Schwarzian theory  \cite{Maldacena:2016upp, kitaev2015simple, Maldacena:2016hyu,Charles:2019tiu,Choi:2021nnq}.}

The BPS energy $W$ \eqref{eqn:FreeEnergy} that is intrinsic to the BPS surface
is computed from \eqref{eqn:Glinfct} by dividing with $T$: 
\begin{align}
    \label{eqn:BPSGibbsPrime}
    W &= \frac{G}{T} = \frac{1}{2}N^2 \left(-\frac{(a+b)(2-a-b)}{6(1-a)(1-b)}\Phi' + 
    \frac{ (a+b)^2(1+a)}{2(1-a)^2(1-b)}\Omega_a ' + \frac{ (a+b)^2(1+b)}{2(1-a)(1-b)^2}\Omega_b ' \right) ~.
\end{align}
The primed potentials
\begin{align}
\label{eqn: primedlmt}
   \Phi' &=\frac{\Phi - 3}{T}   & \Omega'_{a,b} &=  \frac{\Omega_{a,b} - 1}{T}~,
\end{align}
were previously introduced in the general BPS formalism through \eqref{eqn:primPO}.
Depending on the point of view, they are either coordinates in the projective geometry defining the BPS surface or, via the limiting procedure, thermal derivatives evaluated at $T=0$. 

The parameters $a$ and $b$ are complicated functions of $\Phi'$, $\Omega_a'$ and $\Omega_b'$ defined 
by the constraints (\ref{eqn:constraint1}-\ref{eqn:constraint2})
recast in the form:
\begin{align}
    \label{eqn:BPSConsp-1}
    -\frac{\sqrt{a+b+ab}}{1-a}\Omega_a ' -\frac{\sqrt{a+b+ab}}{1-b}\Omega_b ' -\frac{\sqrt{a+b+ab}}{3(a+b)}\Phi' &= \pi~,\\
    \frac{(a+b)(1+a)}{1-a}\Omega_a ' -\frac{(a+b)(1+b)}{1-b}\Omega_b ' +\frac{a-b}{3}\Phi' &= 0~.
    \label{eqn:BPSConsp-2}
\end{align}
Differentiation of the BPS free energy $W$ \eqref{eqn:BPSGibbsPrime} with respect to the primed potentials must give the conserved charges through the thermodynamic relations \eqref{eqn:wderi}. When computing the derivatives, note that, in addition to the explicit dependence of $W$ on the primed potentials, there is implicit dependence through $a, b$ that is given by 
(\ref{eqn:BPSConsp-1}-\ref{eqn:BPSConsp-2}). It is a consistency check on the various formulae that the black hole charges computed this way do in fact agree with (\ref{eqn:BPSQ}-\ref{eqn:BPSJb}). 

\subsubsection{Simplifications of the Constraints}
\label{subsubsec:simconstraints}

In general, it is unpractical to solve the constraints  (\ref{eqn:BPSConsp-1}-\ref{eqn:BPSConsp-2}) and give $a, b$ as explicit functions of the potentials $\Omega'_{a,b}$ and $\Phi'$. However, we can do so in the regime where $1-a \ \sim \  1-b \ \sim \ \epsilon$ is
small and positive. This is a version of the Cardy-like limit \cite{Choi:2018hmj, David:2020ems}. 

In this limit, the constraint \eqref{eqn:BPSConsp-2} shows that the $\Phi'$ term is dominated by the large $\Omega'_{a,b}$ terms. We can partially address this by tuning $\Omega'_a\sim \Omega'_b\sim\epsilon$ but, because $a, b$ are both near unity we have $(a-b)\sim\epsilon$, and so it is insufficient that the $\Omega'_{a,b}$ terms are order unity, they must nearly cancel. Thus:
\begin{equation}
    \frac{\Omega'_a}{1-a} = \frac{\Omega'_b}{1-b}~,
    \label{eqn:cardyeq1}
\end{equation}
up to terms of order $\epsilon$. With this equality, and using $a\sim b\sim 1$ extensively, the other constraint \eqref{eqn:BPSConsp-1} then yields
\begin{equation}
\label{eqn:cardyeq2}
    \frac{\Omega'_a}{1-a} = \frac{\Omega'_b}{1-b}= \frac{\Phi' + 2\pi \sqrt{3}}{12}~.
\end{equation}
This equation is equivalent to presenting $a, b$ as explicit functions of the BPS potentials $\Phi',\Omega'_{a,b}$, as we wanted to do. With this result it is straightforward to also express the 
BPS free energy $W$ in \eqref{eqn:BPSGibbsPrime} as function of these variables: 
\begin{equation}
\label{eqn:CardyW}
    W_{\rm Cardy} =-\frac{N^2}{432}\frac{(\Phi' +2\pi \sqrt{3})^3}{\Omega_a ' \Omega_b '}~.
\end{equation}
This is expression is completely explicit, but it applies only in the Cardy-like limit.

A complementary approach to the awkward constraints (\ref{eqn:BPSConsp-1}-\ref{eqn:BPSConsp-2}) that define $a, b$ as functions of the primed potentials is to implement them using Lagrange multipliers. Introducing two 
multipliers $\Lambda_{1,2}$, we have:
\begin{equation}
\begin{split}
    W &= \frac{1}{2}N^2 \left[ \frac{-(a+b)(2-(a+b))}{6 (1-a)(1-b)}\Phi' +\frac{(a+b)^2(1+a)}{2 (1-a)^2(1-b)}\Omega_a '  +\frac{(a+b)^2(1+b)}{2 (1-a)(1-b)^2}\Omega_b ' \right]\\
    &+ \Lambda_1 \left(-\frac{\sqrt{a+b+ab}}{1-a}\Omega_a ' -\frac{\sqrt{a+b+ab}}{1-b}\Omega_b ' -\frac{\sqrt{a+b+ab}}{3(a+b)}\Phi' -\pi \right) \\
    &+ \Lambda_2 \left(\frac{(a+b)(1+a)}{1-a}\Omega_a ' -\frac{(a+b)(1+b)}{1-b}\Omega_b ' +\frac{a-b}{3}\Phi' \right) ~.
\end{split}
\label{eqn:primedwlag}
\end{equation}
Extremization with respect to $a, b$ give conditions that are solved when:
\begin{align}
    \Lambda_1 &= \frac{1}{2}N^2 \frac{2(a+b)\sqrt{a+b+ab}}{(1-a)(1-b)}~,\\
    \Lambda_2 &= \frac{1}{2}N^2 \frac{-(a-b)}{2(1-a)(1-b)}
    \label{eqn:lambdasols}~,
\end{align}
without imposing any additional relation on $\Phi'$ and $\Omega_{a,b} '$.
Substituting these equations into \eqref{eqn:primedwlag}, we find: 
\begin{equation}
    \begin{split}
        W_{\text{ext}}=-\frac{1}{2}N^2&\Big[\frac{2\pi(a+b)\sqrt{a+b+ab}}{(1-a)(1-b)}-\frac{(a+b)(2a+b+ab)}{(1-a)^2(1-b)}\Omega'_a\\
        &-\frac{(a+b)(a+2b+ab)}{(1-a)(1-b)^2}\Omega'_b-\frac{a+b}{(1-a)(1-b)}\Phi'\Big]~.
    \label{eqn:Wext} 
\end{split}
\end{equation}
Importantly, in this equation $a, b$ are auxiliary variables: $W_{\rm ext}$ is defined only after extremizing $a,b$ with primed potentials kept fixed. 

At this point differentiation with respect to the primed potentials is trivial, there is no need to take implicit dependence via $a, b$ into account, because these auxiliary variables are anyway evaluated at their extremum. Thus, in \eqref{eqn:Wext} the coefficient of each BPS potential $\Phi'$, $\Omega'_{a,b}$ must reproduce its conjugate conserved charge (\ref{eqn:BPSQ}, \ref{eqn:BPSJa},\ref{eqn:BPSJb}), as indeed it does. Additionally, the BPS black hole entropy can be computed from the BPS potential through the formula \eqref{eqn:BPSentro}, which amounts to extracting the independent-of-primed potentials constant in the BPS free energy $W_{\rm ext}$ given in \eqref{eqn:Wext}. It agrees with the BPS entropy \eqref{eqn:S}, as it should. 

The free energy \eqref{eqn:Wext} is thus completely general, 
unlike the expression \eqref{eqn:CardyW} in the Cardy-limit
no assumptions were made on the parameters. The drawback is that, in this formalism, $a, b$ must be extremized over, and the resulting conditions are precisely the constraints (\ref{eqn:BPSConsp-1}-\ref{eqn:BPSConsp-2}).


\subsubsection{Thermodynamics of BPS Black Holes}
\label{subsubsec:BPSthermo1}

In our study of BPS thermodynamics, we will facilitate comparisons with the literature by trading the BPS electric potential $\Phi'$ for
\begin{equation}
\varphi' = \Phi' -\Omega_a ' -\Omega_b '~,
\label{eqn:tildepots}
\end{equation}
which is the obvious BPS analogue of \eqref{eqn: varphidef}.
The constraints (\ref{eqn:constraint1}-\ref{eqn:constraint2}) for the angular BPS potentials
then become:
\begin{align}
\label{eqn:Omegaaphi}
    \Omega_a'=-\frac{1-a}{1+3a+3b+a^2+3ab+b^2}\left[\frac{\pi (a+2b+2ab+b^2)}{\sqrt{a+b+ab}}+\frac{1+a}{2}\varphi'\right]~,\\
\label{eqn:Omegabphi}
    \Omega_b'=-\frac{1-b}{1+3a+3b+a^2+3ab+b^2}\left[\frac{\pi (2a+b+2ab+a^2)}{\sqrt{a+b+ab}}+\frac{1+b}{2}\varphi'\right]~. 
\end{align}
For reference, we also record the inverse formula that converts from $\varphi'$ back to the electric BPS potential $\Phi'$:
\begin{eqnarray}
    \label{eqn:Phip}
    \Phi'=-\frac{3 (a+b)}{1+3a+3b+a^2+3ab+b^2}\left[\frac{\pi (1-ab)}{\sqrt{a+ b+ab} }-\frac{(2+a+b)}{2}\varphi'\right]~.
\end{eqnarray}
In the new variables that eliminate $\Phi'$, the free BPS energy \eqref{eqn:Wext} becomes:
\begin{equation}
    \label{eqn:BPSGibbs}
    \begin{split}
        \frac{W}{N^2}=&\frac{\pi (a+b)^2 \left[1-2a-2b-a^2-5ab-b^2-a^2b-ab^2\right] }{2(1-a)(1-b) \left(1+3a+3b+a^2+3ab+b^2\right)\sqrt{a+b+ab}}\\
        &-\frac{(a+b)^2(3+a+b-ab) }{4 (1-a) (1-b) \left(1+3a+3b+a^2+3ab+b^2\right)}\varphi'~.
    \end{split}
\end{equation}

Even though BPS thermodynamics applies only at strictly vanishing temperature $T=0$, it is meaningful to construct phase diagrams that depend on the BPS potentials. In order to interpret them, it is useful to introduce the ``BPS temperature" 
\begin{align}
\label{eqn:BPST}
    &\tau\equiv-\frac{2}{\Omega_a'+\Omega_b'}=\frac{4 \left(1+3a+3b+a^2+3ab+b^2\right)}{ \frac{6\pi  (a+b)(1-ab)}{\sqrt{a+b+ab}} + \left(2-a^2-b^2\right)\varphi' }~, 
\end{align}
that shares some properties with the usual physical temperature. For example, it is positive because physical black holes have $1 - \Omega_i\geq 0$ and so the primed potentials $\Omega'_i$ introduced in \eqref{eqn: primedlmt} are negative. Also, for small and fixed $1 - \Omega_i$, the effective temperature $\tau$ is proportional to the physical temperature $T$, again by \eqref{eqn: primedlmt}. Finally, it generalizes the analogous definition in \cite{Choi:2018vbz, Choi:2018hmj}.

In the preceding, and in our entire study of the phase diagram for BPS black holes, we have opted to use $(a,b,\varphi')$ as the variables for all thermodynamic quantities. The $a$, $b$ arise as conventional parameters for presenting the underlying BPS black hole geometry, they do not have a meaning directly in thermodynamics. When we write them, we refer to them
as the functions of $\Omega'_i$ and $\varphi'$ that are defined implicitly by the constraints  (\ref{eqn:BPSConsp-1}-\ref{eqn:BPSConsp-2})
which cannot be tractably inverted, except in specific cases like the Cardy-like limit where we have the explicit relation \eqref{eqn:cardyeq2}.


\subsection{The BPS Phase Diagram: a Benchmark Case}
\label{subsec:elemcase}
In this subsection we review the special case where angular momenta are equal ($a=b$) and $\varphi'=0$ \cite{Choi:2018vbz, Choi:2018hmj}. It will serve as an important benchmark for our study of the general black holes in section \ref{sec:BPSthermodynamics}.

\begin{figure}[H]
    \centering
    \includegraphics[width=0.6\textwidth]{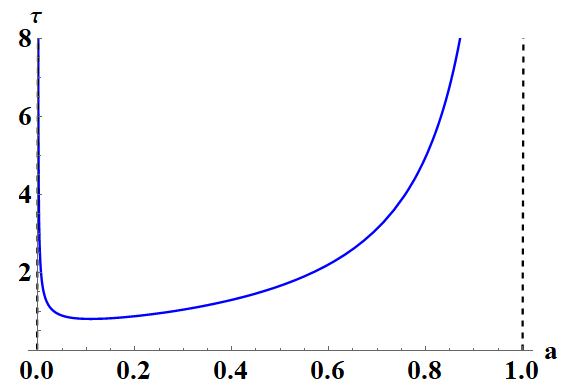}
    \caption{The BPS temperature $\tau$ as function of the rotation parameter $a$ when $a=b$ and $\varphi' = 0$. The local minimum determines the position of the cusp in the phase diagram. The divergences at $a=0$ and $a=1$ correspond to the large-$\tau$ extremes of the ``small" and ``large" black hole branches, respectively.}
    \label{fig:tau-a-02}
\end{figure}

The BPS temperature \eqref{eqn:BPST} simplifies when $a=b$ and $\varphi'=0$:
\begin{align}
    \label{eqn:BPSExampleT}
    &\tau=\frac{1+5 a}{3\pi(1-a)}\sqrt{1+\frac{2}{a}}~.
\end{align}
This expression diverges at $a=0$ and $a=1$. It has a local minimum when the parameter  $a=a_{\text{cusp}}=\frac{3\sqrt{3}-4}{11}\approx0.109$ and $\tau=\tau_{\text{cusp}}\approx0.809$. It is plotted in Figure \ref{fig:tau-a-02}.

The free energy \eqref{eqn:BPSGibbs} when $a=b$ and $\varphi'=0$ is:
\begin{align}
    \label{eqn:BPSExampleW}
      \frac{W}{N^2}=&\frac{2\pi a^{\frac{3}{2}} ( 1- 5a -2a^2)}{(1-a)^2 \left(1 + 5a\right)\sqrt{2+a}}~.
\end{align}
The expression yields $W=0$ for $a=0$ and increases for small and positive $a$. 
The BPS temperature $\tau\to\infty$ as $a\to 0^+$ so in the limit of small parameter $a$ the BPS free energy $W\to 0^+$ for large $\tau$. As $a$ increases to the value $a_{\rm cusp}$ where the temperature is at its minimum $\tau_{\rm cusp}$, the free energy increases monotonically to $W_{\rm cusp}/N^2= \frac{11\pi}{27\sqrt{273+158\sqrt{3}}} \approx 0.055 $.
The range $0<a<a_{\rm cusp}$  maps out the ``small" black hole branch of the BPS phase diagram. 

\begin{figure}[H]
    \centering
    \includegraphics[width=\textwidth]{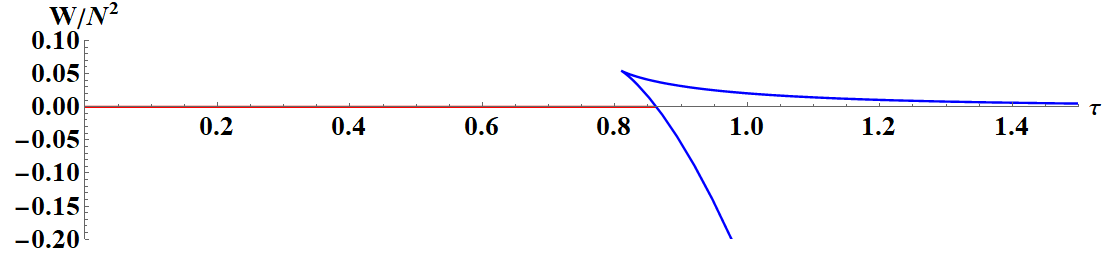}
    \caption{BPS free energy $W/N^2$ vs. BPS temperature $\tau$. This phase diagram displays an upper (small black hole) branch and a lower (large black hole) branch. The two branches coincide at the cusp $(\tau_{\text{cusp}}, W_{\text{cusp}}/N^2) \approx(0.809,0.055)$. The large black hole branch dominates over the BPS gas with $W=0$ (red line) for $\tau > \tau_{\text{HP}}\approx0.863$.}
    \label{fig:W-tau-0}
\end{figure}

As $a$ increases above $a_{\rm cusp}$, the free energy decreases from $W_{\rm cusp}$. It crosses $W=0$ 
at $a_{\text{HP}}=\frac{\sqrt{33}-5}{4}\approx0.186$ which corresponds to the BPS temperature $\tau_{\text{HP}}=0.863$. Finally, in the limit $a\to 1^-$, the free energy diverges $W\to-\infty$ along with the BPS temperature $\tau\to\infty$. The range where $a_{\rm cusp}<a<1$ is the ``large" BPS black hole branch. 
We plot the BPS phase diagram in Figure \ref{fig:W-tau-0}

The phase diagram of the BPS black hole is remarkably similar to that of AdS-Schwarzschild, reviewed in subsection \ref{subsubsec:phi0} \cite{Choi:2018vbz}. In other words, Figure \ref{fig:W-tau-0} is reminiscent of Figure \ref{fig:phase00}. In line with this analogy, we denoted the point where the BPS free energy crosses $W\equiv0$ by the subscript "HP" that refers to the Hawking-Page transition. However, the notion of thermal AdS is not trivial in the BPS context. It presupposes the existence of a non-interacting gas of BPS particles for any $\tau$, corresponding to a general value of the BPS potential for angular momentum $\Omega'$. This hypothetical phase would always have $W=0$ and be preferred for the $\tau$'s where the BPS black hole has $W>0$. The equilibrium conditions between such states of matter have not yet been established so we consider the ``BPS gas" conjectural. 

The striking qualitative agreement between the AdS-BPS and AdS-Schwarzschild phase diagrams hides a more detailed distinction between their asymptotic behaviours. 

Starting with the Schwarzschild-AdS spacetime, the Gibbs free energy $G$ and the temperature $T$ computed from the parameters (\ref{eqn: mass} - \ref{eqn:T}) with $a=b=0$
\begin{align}
\label{eqn:sadsG}
    G &= \frac{N^2r_+^2(1-r_+^2)}{4} ~,\\
\label{eqn:sadsT}
    T &= \frac{1+2 r_+ ^2}{2\pi r_+} ~,
\end{align}
give the rescaled free energy $W= G/T$:
\begin{align}
\label{eqn:sadsW}
    \frac{W}{N^2} &= \frac{\pi }{2} \frac{r_+^3(1-r_+^2)}{1+2 r_+ ^2} ~.
\end{align}
This allows us to extract the asymptotic behaviors of the two branches by focussing on the extreme regimes $r_+\ll 1$ and $r_+ \gg 1$. 

For very small black holes $r_+ \ll 1$, $T$ diverges as $\sim r_+ ^{-1}$ and $W$ vanishes as $\sim r_+^3$
so:
\begin{equation}
    \label{eqn:sadsWasym1}
    W \sim T^{-3} \text{ as } T\to \infty ~. 
\end{equation}
This is equivalent to $M\sim T^{-2}$ for asymptotically flat 5D Schwarzschild black holes. In contrast, 
for very large black holes $r_+ \to \infty$, $T$ diverges as $r_+$ and $W$ diverges as $-r_+ ^3$
so then:
\begin{equation}
\label{eqn:sadsWasym2}
W \sim -T^{3} \text{ ~~as ~~} T\to \infty ~.
\end{equation}
as expected for a conformal fluid in the dual CFT$_4$. 

Turning now to the BPS black holes, we use equations (\ref{eqn:BPSExampleT}-\ref{eqn:BPSExampleW}). 
For very small BPS black holes $a \to 0^+$, $\tau$ diverges $\sim a^{-1/2}$ and $W$ vanishes as $\sim a^{3/2}$ so:
\begin{equation}
\label{eqn:BPSWasym1}
W \sim \tau^{-3} \text{ ~~as~~ } \tau\to \infty ~. 
\end{equation}
When $a \to 1^-$ instead, $\tau$ diverges as $\sim (1-a)^{-1}$ and $W$ diverges as $\sim -(1-a)^{-2}$ so:
\begin{equation}
\label{eqn:BPSWasym2}
W \sim -\tau^{2} \text{ ~~as ~~} \tau \to \infty ~.
\end{equation}

Our results show that the asymptotic behaviors (\ref{eqn:sadsWasym1}, \ref{eqn:BPSWasym1}) for very small black holes agree but the analogues (\ref{eqn:sadsWasym2}, \ref{eqn:BPSWasym2}) 
for large black holes do not. Thus the qualitative identification between $\tau $ and $T$ is not precise.

\subsection{The HHZ Free Energy}
\label{subsubsec: HHZpota=b}

The HHZ free energy has been central to recent progress on the microscopic understanding of BPS black holes in AdS \cite{Benini:2015noa, Benini:2015eyy, Benini:2016rke, Hosseini:2017mds, Cabo-Bizet:2018ehj, Choi:2018hmj, Benini:2018ywd, Choi:2019miv, Zaffaroni:2019dhb}. It was originally motivated by analysis of the near-horizon geometry of supersymmetric black holes rather than conventional black hole thermodynamics. In this subsection we address this gap in the literature.

\subsubsection{Extremization of the HHZ Free Energy}
\label{subsubsec:hhzextremize}

The HHZ free energy is a functional of the HHZ potentials $\Delta$ and $\omega_{a,b}$ that are conjugate to the conserved charges $Q$ and $J_{a,b}$: 
\footnote{Our reference to the HHZ free energy generalizes the original research \cite{Hosseini:2017mds}. In particular, we allow general, non-equal angular momenta $J_{a,b}$.}
\begin{equation}
\label{eqn:hhzpot}
    H (\Delta, \omega_{a,b}) = -\frac{N^2}{2} \frac{\Delta^3}{\omega_a \omega_b}~.
\end{equation}
Unlike the free energy that is familiar from undergraduate studies, the HHZ free energy is complex and the potentials it depends on are complex as well. The latter satisfy the constraint
\begin{equation}
    \label{eqn:BPSDeltaomegacond}
    3\Delta - \omega_a - \omega_b = 2\pi i~, 
\end{equation}
so they are genuinely complex numbers \cite{Hosseini:2017mds, Zaffaroni:2019dhb}. 

In order to extremize the free energy \eqref{eqn:hhzpot} over its arguments $\Delta$, $\omega_{a,b}$, while taking into account the constraint \eqref{eqn:BPSDeltaomegacond}, it is convenient to Legendre transform to an ensemble with fixed charges $Q, J_{a,b}$. 
To do so, we introduce the 
 entropy function: 
\begin{equation}
\label{eqn:entropyfunc}
    S(\Delta, \omega_{a,b}, Q, J_{a,b}) = -\frac{N^2}{2} \frac{\Delta^3}{\omega_a \omega_b} - 3\Delta Q - \omega_a J_a - \omega_b J_b  -\Lambda (3\Delta -\omega_a - \omega_b - 2\pi i)~,
\end{equation}
where $\Lambda$ is a Lagrange multiplier. 
Extremization over the potentials then gives:
\begin{align}
\label{extremeq1}
    \frac{\partial S}{\partial \Delta} = 0 & ~~~\implies~~~ -\frac{N^2}{2}\frac{3\Delta^2}{\omega_a \omega_b} -3(Q+\Lambda) = 0 ~, \\
\label{extremeq2}
    \frac{\partial S}{\partial \omega_i} = 0 & ~~~\implies~~~ \frac{N^2}{2\omega_i}\frac{\Delta^3}{\omega_a \omega_b} -(J_i-\Lambda) = 0 ~.
\end{align}
When these conditions are satisfied, the entropy at the extremum $S^*$ is related to the Lagrange multiplier $\Lambda$ through
\begin{equation}
    S^* = 2\pi i \Lambda~,
    \label{eqn:sstrlm}
\end{equation}
and $\Lambda$ is determined by the black hole charges
$Q$, $J_{a,b}$ via the cubic equation: 
\begin{equation}
\label{eqn:Lambdaeq}
    \Lambda^3 + \underbrace{(3Q+\frac{N^2}{2})}_{A} \Lambda^2 +\underbrace{ \left(3Q^2 -\frac{N^2}{2} (J_a + J_b) \right)}_{B} \Lambda + \underbrace{(Q^3 +\frac{N^2}{2}J_a J_b)}_{C} = 0 ~.
\end{equation}
The HHZ prescription invokes reality and positivity of the entropy, so that $\Lambda$ must be a purely imaginary number with negative imaginary part. This is possible only if the cubic polynomial factorizes as $(\Lambda^2 + B)(\Lambda+A)$. It follows that
the coefficients of the cubic polynomial satisfy the relation:
\begin{equation}
\label{eqn:chargeconst}
    AB-C = (3Q+\frac{N^2}{2})\left(3Q^2 -\frac{N^2}{2} (J_a + J_b) \right) - (Q^3 +\frac{N^2}{2}J_a J_b) = 0~,
\end{equation}
and also that the root with negative purely imaginary part is:
\begin{equation}
    \Lambda = - i \sqrt{B} = -i\sqrt{3Q^2 -\frac{N^2}{2} (J_a + J_b)}~.
    \label{eqn:defLambda}
\end{equation}
Inserting this value for $\Lambda$ in \eqref{eqn:sstrlm}, and expressing $Q$ and $J_{a,b}$ in terms of $a$ and $b$ using their values from (\ref{eqn:BPSQ}-\ref{eqn:BPSJb}), we match the BPS black hole entropy $S^*$ \eqref{eqn:BPSS}. 
Moreover, the condition \eqref{eqn:chargeconst} is precisely the constraint on black hole charges \eqref{eqn:chargeconstraint}. 

The HHZ procedure thus gives the correct functions of charges, but we stress that we introduced the HHZ free energy $H$ \eqref{eqn:hhzpo} and the potentials $(\Delta, \omega_{a,b})$ it depends on without committing to an associated spacetime interpretation: we derived the black hole entropy via a formal extremization procedure. Some authors simply identify the HHZ formalism with ``gravitational thermodynamics", and consider it the target for microscopic CFT considerations. However, the HHZ variables are genuinely complex so their relation to spacetime physics is not clear {\it a priori}. We therefore distinguish ``boundary CFT" (the HHZ free energy) and ``bulk interpretation" (black hole thermodynamics) carefully.

The values of the potentials $\Delta$ and $\omega_i$ (with $i=a, b$) that extremize the HHZ free energy follow from (\ref{extremeq1}-\ref{extremeq2})
as:
\begin{align}
\label{eqn:deltaext}
    \Delta &= 2\pi i(Q+\Lambda)^{-1} \left(\frac{1}{J_a - \Lambda} + \frac{1}{J_b - \Lambda} + \frac{3}{Q+\Lambda} \right)^{-1}~, \\ 
    \omega_i &= -2\pi i(J_i - \Lambda)^{-1} \left(\frac{1}{J_a - \Lambda} + \frac{1}{J_b - \Lambda} + \frac{3}{Q+\Lambda} \right)^{-1}~.
    \label{eqn:omegaext}
\end{align}
with the understanding that the Lagrange multiplier $\Lambda = \Lambda(Q,J_{a,b})$ is given by \eqref{eqn:defLambda}. 
Because of the constraint  \eqref{eqn:BPSDeltaomegacond}, they yield the on-shell value of the HHZ free energy \eqref{eqn:hhzpot}
\begin{equation}
H = -\frac{N^2}{2} \frac{1}{\omega_a \omega_b}\left(\frac{2\pi i+\omega_a + \omega_b}{3}\right)^3 ~.
\label{eqn:hhzpo}
\end{equation} 
    
The values of the potentials $\Delta,\omega_i$ (\ref{eqn:deltaext}-\ref{eqn:omegaext}) and the entropy $S$ 
\eqref{eqn:sstrlm} at the extremum are all given in terms of the charges $Q, J_i$. This is a natural form to present the result of the extremization, and the one that is most appropriate for comparison with microscopic considerations. However, it is awkward that all these expressions are defined only modulo the constraint on the charges \eqref{eqn:chargeconst}.

The gravitational thermodynamics of BPS black holes expresses the corresponding physical quantities in terms of the parameters $(a, b)$, in a form that automatically solves the constraint on the charges. 
In particular, the BPS values of the charge $Q^*$ \eqref{eqn:BPSQ} and angular momenta $J_i ^*$ (\ref{eqn:BPSJa}-\ref{eqn:BPSJb}) satisfy the constraint and give equations for $\omega_i$:
\begin{align}
\label{eqn:omegaai}
    \omega_a &= \frac{-\pi (1-a)}{a^2 +b^2 +3(a+b+ab)+1} \left(\frac{a+2b+b^2+2ab}{\sqrt{a+b+ab
    }} + i (1+a)\right)~, \\
\label{eqn:omegabi}
    \omega_b &= \frac{-\pi (1-b)}{a^2 +b^2 +3(a+b+ab)+1} \left(\frac{b+2a+a^2+2ab}{\sqrt{a+b+ab
    }} + i (1+b)\right)~.
\end{align}
We do not explicitly write the corresponding expression for the electric potential $\Delta= \Delta(a,b)$ because it follows from the simple linear relation \eqref{eqn:BPSDeltaomegacond}. The on-shell value of the HHZ potential $H$ \eqref{eqn:hhzpo}
can be similarly expressed in terms of $a$ and $b$:
\begin{equation}
\begin{split}
\label{eqn:onshellH}
    H(a,b) &= \frac{N^2}{2} \frac{-\pi (a+b)^2 (1-2a-2b-a^2-5ab-b^2-a^2 b-a b^2)}{(1-a)(1-b)\sqrt{a+b+ab} (a^2+b^2+3(a+b+ab)+1)} \\
    &+ \frac{N^2}{2} \frac{\pi(a+b)^2(3-ab+a+b)}{2(1-a)(1-b)(a^2+b^2+3(a+b+ab)+1)}i~.
\end{split}
\end{equation}

Even though the potentials $\omega_{a,b}$ in  (\ref{eqn:omegaai}-\ref{eqn:omegabi}) and the HHZ potential \eqref{eqn:onshellH} are functions of the parameters $a, b$ that were introduced in the context of gravitational solutions, we stress that these formulae are entirely equivalent to (\ref{eqn:deltaext}-\ref{eqn:hhzpo}) which we consider function of charges. The only difference is that the constraint on charges was solved. As a bonus, the potentials that extremize the HHZ free energy are now written in a form that is convenient for comparison with gravitational thermodynamics.  

\subsubsection{Comparing the HHZ Results with Black Hole Thermodynamics}
\label{subsec:gravresults}

We now reconsider the conventional thermodynamics of BPS black holes that depends on manifestly real spacetime potentials $\Phi', \Omega'_{a,b}$ and the BPS free energy $W$  \eqref{eqn:BPSGibbs} that is also real. Our gravitational considerations
presented results in terms of auxiliary variables $(a, b)$ that are related to black hole charges and an additional physical potential $\varphi'=\Phi'-\Omega'_{a}-\Omega'_{b}$ introduced in \eqref{eqn:tildepots}.
In these variables, the BPS free energy $W$ defined in \eqref{eqn:BPSGibbsPrime} becomes \eqref{eqn:BPSGibbs}:
\begin{equation}
    \begin{split}
        \frac{W}{N^2}=&\frac{\pi (a+b)^2 \left(1-2a-2b-a^2-5ab-b^2-a^2b-ab^2\right) }{2(1-a)(1-b) \left(1+3a+3b+a^2+3ab+b^2\right)\sqrt{a+b+ab}}\\
        &-\frac{(a+b)^2(3+a+b-ab) }{4 (1-a) (1-b) \left(1+3a+3b+a^2+3ab+b^2\right)}\varphi'~~,
    \end{split}
\end{equation}
and the BPS angular velocities defined in \eqref{eqn: primedlmt} are:

\begin{align}
\label{eqn:OmegaPa}
        \Omega_a' &=-\frac{1-a}{2(1+3a+3b+a^2+3ab+b^2)}\left[\frac{2\pi (a+2b+2ab+b^2)}{\sqrt{a+b+ab}}+(1+a)\varphi'\right]~,\\
    \label{eqn:OmegaPb}
    \Omega_b'&=-\frac{1-b}{2(1+3a+3b+a^2+3ab+b^2)}\left[\frac{2\pi (2a+b+2ab+a^2)}{\sqrt{a+b+ab}}+(1+b)\varphi'\right]~.
\end{align}
The BPS electric potential $\Phi'$ defined in \eqref{eqn: primedlmt} is not an independent variable because it is given through 
$\Phi' = \Omega'_{a}+\Omega'_{b} + \varphi'$. 

We expect that the spacetime free energy $W$  \eqref{eqn:BPSGibbs} must be related to the HHZ free energy $H$ \eqref{eqn:onshellH}. Similarly, the spacetime angular velocities $\Omega'_{a,b}$ (\ref{eqn:Omegaaphi}-\ref{eqn:Omegabphi}) should
be related to the HHZ variables $\omega_{a,b}$ (\ref{eqn:omegaai}-\ref{eqn:omegabi}), since they are both conjugate to angular momenta $J_{a,b}$. There are indeed striking similarities between the expressions but there are also key differences. As we have stressed already, the HHZ variables are complex while black hole thermodynamics is manifestly real. Moreover, the thermodynamic formulae depend on one more variable $\varphi'$. 

There are several options that address these differences, at least formally. The most conservative is to assign $\varphi'$ a specific value that is imaginary. Indeed, for $\varphi' = 2\pi i$ we have:
\begin{equation}
\label{eqn:HWrelation}
    H(a,b) = - W(a,b,\varphi' = 2\pi i)~, 
\end{equation}
for the free energy and 
\begin{equation}
\label{eqn:omegaDeltaHHZmatch}
 \omega_i (a,b) = \Omega_i ' (a,b,\varphi' = 2\pi i) ~~~,~    3\Delta(a,b) = \Phi'(a,b,\varphi'=2\pi i) ~,
\end{equation}
for the potentials. Moreover, the definition of $\varphi'$ reduces precisely to the HHZ constraint \eqref{eqn:BPSDeltaomegacond} when $\varphi'=2\pi i$.

This procedure is well motivated. The general BPS partition function reduces to the superconformal index precisely for $\varphi'=2\pi i$ because $\varphi'$ is integral (half-integral) for bosons (fermions), and so this value of the potential is equivalent to inserting $(-)^F$. Therefore, for this value of $\varphi'$, comparisons between weak and strong coupling in the CFT is justified, with the latter dual to the semiclassical gravity description. 
According to this line of reasoning, black hole thermodynamics for any real value of $\varphi'$ corresponds to strongly coupled field theory that is unlikely to be accounted for by computations in weakly coupled CFT. 

A distinct procedure, also common in the literature \cite{Hosseini:2016tor, Liu:2017vbl, Choi:2018vbz, Choi:2018hmj, Larsen:2019oll, Copetti:2020dil}, is to identify only the real parts of the HHZ free energy and potentials as physical, and compare them with black hole thermodynamics only for the special value $\varphi'=0$. Indeed, for the free energy, we find the identity:
\begin{equation}
\label{eqn:reHab}
    \text{Re} \ H(a,b) = - W(a,b,\varphi'=0) = \frac{N^2}{2} \frac{-\pi(a+b)^2(1-2a-2b-a^2-5ab-b^2-a^2 b-a b^2)}{(1-a)(1-b)\sqrt{a+b+ab}(a^2+b^2+3(a+b+ab)+1)}~.
\end{equation}
Sometimes authors even refer to the quantity $\text{Re} \ H(a,b)$ (expressed in terms of the HHZ variables $\Delta$ and $\omega_{a,b}$) as the free energy $F$, rather than the full complex expression \cite{Choi:2018vbz, Choi:2018hmj, Larsen:2019oll, Copetti:2020dil}. The analogous comparison between the real part of the complex chemical potentials is also successful:
\begin{equation}
\label{eqn:omegaOmegarels}
    \begin{split}
        \text{Re } \omega_{a} &=  \frac{-\pi (1-a)(a+2b+b^2+2ab)}{(a^2 +b^2 +3(a+b+ab)+1)\sqrt{a+b+ab}} = \Omega_a ' (a,b,\varphi'=0)~,  \\
        \text{Re } \omega_{b}&=  \frac{-\pi (1-b)(2a+b+a^2+2ab)}{(a^2 +b^2 +3(a+b+ab)+1)\sqrt{a+b+ab}} = \Omega_b ' (a,b,\varphi'=0) ~.
    \end{split}
\end{equation}
The focus on the real part of the potential is heuristic, but it is striking that it ``works", in that it yields identities between somewhat elaborate functions. Within the rigorous framework of the superconformal index, these agreements are 
coincidental.

The feature that allows {\it both} procedures to work is that the free energy is {\it linear} in  $\varphi'$. The analogous feature in the AdS$_3$/CFT$_2$ correspondence has an appealing physical interpretation: the BPS black hole cannot be identified with a specific BPS state, it is an average over all chiral 
primaries \cite{Larsen:2021wnu}. It would be interesting to develop an analogus interpreation in the AdS$_5$/CFT$_4$ correspondence. 

\subsubsection{HHZ free Energy and Black Hole Thermodynamics: the Phase Diagram}
\label{subsubsec:HHZChoi}

 Choi {\it et. al.} showed that a phase diagram determined from the HHZ free energy \eqref{eqn:hhzpot} shares important aspects of the AdS-Schwarzschild black  thermodynamics, including the Hawking-Page phase transition and its characteristic ``cusp" \cite{Chamblin:1999tk, Witten:1998qj, Natsuume:2014sfa, Kubiznak:2016qmn}. This is interesting, because it suggests that the confinement/deconfinement transition in QCD-like theories \cite{tHooft:1977nqb, Polyakov:1978vu, Witten:1998zw, Susskind:1979up,Aharony:2003sx} can be analyzed while preserving supersymmetry, possibly with a great deal of precision. 
However, it is not obvious that results derived from the HHZ free energy agree with traditional thermodynamics, derived from black hole geometry. In this subsection we verify that they do in fact agree. 

\begin{figure}[H]
    \centering
    \includegraphics[width=0.6\textwidth]{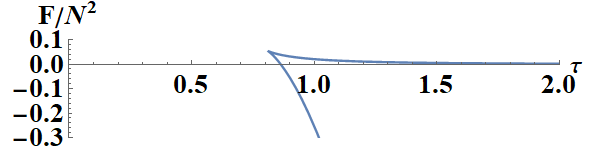}
    \caption{The free energy $F$ vs. the BPS temperature $\tau$. This phase diagram was derived from the HHZ potential using the ``heuristic" method.}
    \label{fig:wtau-choi}
\end{figure}

The extremization conditions on the entropy function (\ref{extremeq1},\ref{extremeq2}) give:
\begin{equation}
\label{eqn:JpQomega}
    J+Q = - \frac{2N^2}{27}\frac{(\omega + \pi i)^2(\omega-2\pi i)}{\omega^3}~,
\end{equation}
after taking $\omega_a=\omega_b$ and eliminating $\Lambda$. Choi {\it et.al.} follow the ``heuristic" procedure yielding (\ref{eqn:reHab}-\ref{eqn:omegaOmegarels}) so they demand that $J, Q$ are positive and real. This requirement relates the real and imaginary parts of $\omega=\omega_R + i \omega_I$ as: 
\begin{equation}
    \label{eqn:omegarirelation}
    \omega_R^2 = \frac{3\omega_I^2(\pi + \omega_I)}{\pi - 3\omega_I}~.
\end{equation}
The heuristic procedure interprets $\omega_R = \text{Re }\omega$ as the
physical potential $\Omega'$ through \eqref{eqn:omegaOmegarels} and the condition that $\Omega'<0$ determines phases so: 
\begin{equation}
        \label{eqn:choitau}
        \tau = -\omega_R^{-1} = 
        -\omega_I^{-1} \sqrt{\frac{\pi - 3\omega_I}{3(\pi + \omega_I)}}~,
\end{equation}
and constrains the imaginary part so $-\pi<\omega_I<0$. 
Further, the heuristic procedure posits that it is the real part of the HHZ energy \eqref{eqn:hhzpot} that is physical. Since $\Delta$ is related to $\omega=\omega_R + i \omega_I$ through the constraint \eqref{eqn:BPSDeltaomegacond} we then find
\begin{align}
\label{eqn:choiF}
    F &= -\frac{N^2}{2} \text{Re } \frac{\Delta^3}{\omega^2} = -\frac{N^2}{2} \text{Re } \left(\frac{(2\pi i + 2\omega)^3 }{27\omega^2}\right)\\
    &= -\frac{N^2}{18}\frac{\pi^3 - 9\pi \omega_I ^2 -8\omega_I^3}{\omega_I ^2}\sqrt{\frac{\pi + \omega_I}{3(\pi -3\omega_I)}}~.
\end{align}
The variable $\omega_I$ is related by the BPS temperature $\tau$ through 
\eqref{eqn:choitau} so this equation gives the free energy $F$ as function of the BPS temperature $\tau$ which we
plot in Figure \ref{fig:wtau-choi}. 

From visual inspection it seems clear that Figure \ref{fig:wtau-choi} is precisely the same as its analogue in Figure \ref{fig:W-tau-0} that applies to black holes with parameters $a=b$ and $\varphi'=0$. However, they were arrived at very differently: the former by extremizing the HHZ potential \eqref{eqn:hhzpot} (and imposing various reality conditions), the latter from the standard thermodynamic interpretation of BPS black hole geometry. 

The analytical comparison is also quite nontrivial. The BPS free energy $W$ derived from gravity is a function of the BPS temperature $\tau$, with both $W$ and $\tau$ given in terms of the  auxiliary parameter $a$ in (\ref{eqn:BPSExampleT},\ref{eqn:BPSExampleW}), and reproduced here for convenience: 
\begin{align}
    W &= \frac{N^2}{2} \frac{4\pi a^2(1-5a-2a^2)}{(1-a)^2(1+5a)\sqrt{a(a+2)}} ~ , \\
    \tau & = -\omega^{-1}_R = -\frac{3\pi a(1-a)}{(1+5a)\sqrt{a(a+2)}} ~ .
\end{align}
It must be compared with the real part of the HHZ free energy $F$ \eqref{eqn:choiF} that is a function of $\omega_I$, the imaginary part of the complexified rotational velocity $\omega=\omega_R + i \omega_I$, which serves as an auxiliary parameter that is related to $\tau=-\omega_R^{-1}$ through \eqref{eqn:choitau}. 

As suspected, there is in fact an identification of auxiliary parameters that transforms all the analytical formulae into one another: 
\begin{equation}
\label{eqn:witoarel}
    \omega_I  = - \pi \frac{1-a}{1+5a} ~. 
\end{equation}
We stress that, given the nonlinear nature of the formulae involved, this agreement was far from preordained. Its success is a sensitive test of the HHZ potential \eqref{eqn:hhzpot}.

\section{Thermodynamics of BPS Black Holes: Detailed Study}
\label{sec:BPSthermodynamics}

In this section we study the phase diagram of general BPS black holes. 
The benchmark BPS black hole introduced in subsection \ref{subsec:elemcase} has equal angular velocities $\Omega'_a=\Omega'_b$ and electric potential $\Phi'$ tuned 
so $\varphi'=0$. We now explore generic values of the three independent potentials and explain their significance. \\

\subsection{Primed Potentials and Their Conjugate Charges}
\label{subsec:chargepotsconjug}

The general BPS black holes we consider in this section are parametrized by the three
primed potentials $\Omega'_a$, $\Omega'_b$, and $\Phi'$. The primes 
remind us that their definitions \eqref{eqn: primedlmt} relate them to the corresponding 
non-BPS potentials $\Omega_{a,b}$ and $\Phi$ via a thermal derivative. The primed potentials are the variables that the BPS free energy $W$
\begin{equation}
    \label{eqn:FreeEnergyvarphi}
    W = -S - \Omega'_a J_a  - \Omega'_b J_b - \Phi' Q ~,
\end{equation}
defined in \eqref{eqn:FreeEnergy} depends on. In particular, its derivatives read off from the first law of BPS black hole thermodynamics 
\begin{equation}
    \label{eqn:FreeEnergyvarphid}
    dW = -  J_a d\Omega'_a  -  J_b d\Omega'_b -  Q d\Phi'~,
\end{equation}
yield the black hole charges. Alternative ensembles that are functions of some or all of the charges, rather than their conjugate potentials, can be obtained as usual, by appropriate Legendre transforms. 

As we discussed in subsection \ref{subsubsec:BPSthermo1}, the interpretation of the phase diagram is simplified by introducing a different basis for the potentials. The BPS temperature $\tau$ \eqref{eqn:BPST} is an analogue of the physical temperature. The modified potential 
$\varphi'$ \eqref{eqn:tildepots} is a proxy for the electric field that is defined so the benchmark case, studied in the literature and reviewed in subsection \ref{subsec:elemcase}, corresponds to $\varphi'=0$. To complete a well-defined transformation from the three potentials $\Omega'_{a,b}, \Phi'$ we introduce a third potential:
\begin{equation}
    \label{eqn:mudef}
    \mu = \Omega'_a - \Omega'_b ~,
\end{equation}
that is sensitive to departures from two equal angular momenta. The combinations of charges that are conjugate to the potentials $(\tau, \varphi', \mu)$ follow from the first law of BPS black hole thermodynamics in the form
\begin{equation}
    dW = - \frac{1}{\tau^2}(J_a + J_b + 2Q)d\tau   -  Q d \varphi' -\frac{1}{2}(J_a - J_b)d\mu ~. 
    \label{eqn: firstlaw2}
\end{equation}
For completeness, we also record the inverse transform from the variables we employ to discuss the phase diagram to the original physical BPS potentials:
\begin{align}
\label{eqn:potstransform}
    \Phi' &= -\frac{2}{\tau} + \varphi' ~,\\
    \Omega'_a &= -\frac{1}{\tau} + \frac{\mu}{2} ~, \\ 
    \Omega'_b &= -\frac{1}{\tau} -\frac{\mu}{2} ~,
\end{align}
that are conjugate the physical charges $Q, J_a, J_b$.

The changes of variables above are routine, at the face of it: BPS black holes are characterized by three charges charges $\{Q, J_a, J_b\}$ which, in the grand canonical ensemble, correspond to three BPS potentials $\{\Phi', \Omega'_a , \Omega'_b\}$. When presenting explicit phase diagrams we further change the basis among the potentials to $\{\tau, \varphi', \mu\}$, in order to clarify physical interpretations and connections to the literature. However, straightforward as these transformations appear, they do not incorporate the fact that all BPS black holes satisfy the constraint between charges \eqref{eqn:chargeconstraint}. Mathematically, this means the Legendre transform, from the microcanonical to the canonical ensemble, is singular. 

Our physical interpretation of this peculiar feature, discussed at length in the introduction, is that all classical BPS black holes correspond to thermal equilibrium along the direction in parameter space that corresponds to the BPS potential $\varphi'$. The formula for the mass excess $M-M_*$ \eqref{eqn:MMstarTP} offers a perspective. BPS black holes $M=M_*$ must have zero temperature $T=0$ as well as the potential $\varphi=0$ but the BPS condition does not specify the ``slope" $\varphi'=\varphi/T$ when the approach is realized as a physical limit where $\varphi\to 0$ and $T\to 0$ simultaneously. This ratio is physical: any supersymmetry implies $M-M^*=0$ but, once a given supercharge has been committed to, others that differ by a phase $\varphi'$ are inconsistent with the ``preferred" supersymmetry. \footnote{Slowly varying fluctuations of $\varphi'$ can be identified with the scalar that carries R-symmetry charge in the ${\cal N}=2$ Schwarzian theory describing excitations in the near geometry of the BPS black hole\cite{Fu:2016vas,Iliesiu:2020qvm,Heydeman:2020hhw,Choi:2021nnq} (and references therein).}

The equilibrium condition along the $\varphi'$ direction was understood in AdS$_3$, where it corresponds to black holes necessarily taking on the R-charge that is the average of all chiral primaries, in contrast to the index that is independent of the charges of these states \cite{Dijkgraaf:2000fq,Larsen:2021wnu}. It is important to elucidate the analogous mechanism in AdS$_5$ so, in the following subsections, we treat $\varphi'$ as an independent thermodynamic variable. In particular, we do not take $\varphi'=0$ {\it a priori} because, even if some argument were to establish $\varphi'=0$ as the equilibrium value, we would still need to show that the constraint among charges emerges as a derivative with respect to this variable.

\subsection{The Physical Range of $\varphi'$ and $\mu$}
\label{subsec:genvarphimu}

The BPS free energy $W$ was defined in subsection \ref{subsec:genBPSthermo} as a function of the BPS potentials $\Phi',\Omega'_{a,b}$ but, as have discussed above (and in subsection \ref{subsubsec:BPSthermo1}), the variables $(\tau, \mu, \varphi')$ are preferable. Unfortunately, in practice the free energy and the various potentials are known only in parametric forms, as functions of $(a, b, \varphi')$. For example, the BPS temperature $\tau$ was presented in 
\eqref{eqn:BPST} and (\ref{eqn:OmegaPa}-\ref{eqn:OmegaPb}) similarly give
\begin{equation}
\label{eqn:BPSmu}
    \mu=\frac{a-b}{2(1+3a+3b+a^2+3ab+b^2)}\left[\frac{2\pi(1+2a+2b+ab) }{\sqrt{a+b+ab} }+(a+b)\varphi'\right]~.
\end{equation}

We need to determine the physical regime of the various parameters. The rotational velocities are bounded by the speed of light $\Omega_i\leq 1$ which, via the definition of primed potentials \eqref{eqn: primedlmt}, corresponds to $\Omega'_i <0$. This in turn implies that $\tau$ introduced in \eqref{eqn:BPST} is strictly positive $\tau>0$ so
\begin{align}
\label{eqn:minvarineq}
- \varphi' \leq  \frac{6\pi(a+b)(1-ab)}{(2-a^2-b^2)\sqrt{a+b+ab}}  ~.
\end{align}
Additionally, we require $0\leq a, b<1$, in order that the underlying black hole solutions exist as regular geometries. 

The positive temperature condition in the form \eqref{eqn:minvarineq} is satisfied for all $\varphi' \geq 0$ but it is nontrivial for $\varphi'<0$. For $a=b$, it reduces to 
\begin{equation*}
\label{eqn:minvaraeqb}
-\varphi'  \underset{a=b}{\leq } 6\pi\sqrt{\frac{a}{2+a}} < 2\sqrt{3}\pi ~. 
\end{equation*}
The middle formula is a monotonically increasing function of $a$ and the second inequality would be saturated for $a=1$. Therefore, for equal angular momenta $a=b$, the BPS potential $\varphi'$ is bounded from below by $\varphi_{-}'=-2\sqrt{3}\pi < \varphi' $

For given $a+b$, the function on the right hand side of \eqref{eqn:minvarineq} increases monotonically as function of $(a-b)^2$. Therefore, the lower bound on $\varphi'$ is relaxed when the two angular momenta are unequal. Its global minimum $\varphi'_{\text{min}}=-6\pi$ corresponds to $(a,b) = (1,0)$ or $(a,b) = (0,1)$.

\begin{figure}[H]
    \centering
    \includegraphics[width=0.5\textwidth]{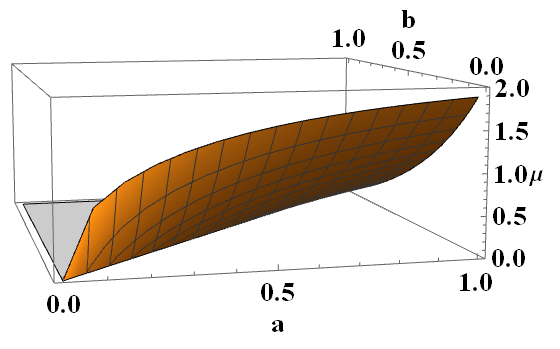}
    \caption{The chemical potential $\mu$ as a function of the black hole parameters $a$ and $b$, for $\varphi'=0$. The maximum is achieved at $(a, b)=(1,0)$. We plot only $a\geq b$, the mirror for $a\leq b$ (the grey area) follows from antisymmetry of $\mu(a,b)$ under $a\leftrightarrow b$.}
    \label{fig:mu-a-b-0}
\end{figure}

We want to similarly analyze the physical range of $\mu$ \eqref{eqn:BPSmu}, the BPS potential that measures the departure from angular momenta of $a$ and $b$ type being equal. Because of the antisymmetry between $a$ and $b$ we have $\mu\equiv0$ along the line $a=b$ and it is sufficient to study $a\geq b$. When $\varphi'=0$ we find that $\mu$ increases monotonically with $a-b$ when $a+b$ is fixed. The
maximal value $\mu_{\rm max}= \frac{3\pi}{5} \approx 1.88$ is reached when $(a,b) = (1,0)$. We plot the BPS potential $\mu$ in Figure \ref{fig:mu-a-b-0}.

\subsection{BPS Black Holes with General $\varphi'$ and Equal Angular Momenta}
\label{subsec:BPSvphimu0}

In this subsection we study the significance of the potential $\varphi'$. We take equal angular momenta, i.e. $a=b$ (and so $\Omega_a'=\Omega_b'=\Omega'$) which corresponds to $\mu\equiv0$.

The BPS free energy \eqref{eqn:BPSGibbs} reduces to
\begin{equation}
    \frac{W}{N^2}=\frac{2a}{(1-a)^2(1+5a)}\left(\frac{\pi(1-5a-2a^2)}{\sqrt{1+\frac{2}{a}}}-\frac{a(3-a)\varphi'}{2}\right)~,
    \label{eqn:BPSGa=b}
\end{equation}
and the BPS ``temperature" $\tau$ \eqref{eqn:BPST} becomes
\begin{equation}
    \tau\equiv-\Omega'^{-1}=\frac{1+5 a}{1-a}\frac{1}{\frac{3 \pi}{\sqrt{1+\frac{2}{a}}}+\frac{\varphi'}{2}}~.
    \label{eqn:BPSTa=b}
\end{equation}
We will discuss the dependence of $W$ on $\tau$ for each sign sign of $\varphi'$ in turn. 
 
\subsubsection{$\varphi'\leq0$}
\label{subsubsec:negativevarphi}

We first consider the BPS temperature \eqref{eqn:BPSTa=b}. For large black holes $\tau$ diverges at $a=1$ and, as $a$ get smaller, it decreases to a minimal temperature $\tau_{\rm cusp}$ that is attained at $a_{\rm cusp}$. This is qualitatively similar to the benchmark case $\varphi'=0$ that was discussed in subsection \ref{subsec:elemcase} (with $\tau(a)$ plotted in figure \ref{fig:tau-a-02}). 

The temperature increases again when $a$ decreases below $a_{\rm cusp}$, as expected for small black holes. However, when $\varphi'<0$, the denominator in the second factor of $\tau$ given in \eqref{eqn:BPSTa=b} reaches zero at some positive value
\begin{equation}
\label{eqn:amin}
a_{\min}=\frac{2\varphi'^2}{36\pi^2-\varphi'^2}~,
\end{equation}
and then the temperature diverges. Thus, when $\varphi'\leq0$, the parameter $a$ is limited to the range $1>a>a_{\min}$.

The parameter value $a_{min}$ increases monotonically when the absolute value $|\varphi'|$ increases. The physical range $1>a>a_{\rm min}$ shrinks to zero if it reaches $a_\text{min}=1$ and then no underlying black hole geometry would exist. Therefore, the corresponding value of the potential $\varphi'$ 
\begin{equation}
\label{eqn:varplim}
\varphi'_{-}=-2\sqrt{3}\pi~,
\end{equation}
constitutes a lower limit $\varphi'\geq \varphi'_{-}$. 

The finite range of $a$ changes the BPS free energy $W$ \eqref{eqn:BPSGa=b} qualitatively: at the high temperature end of the small black hole branch (as $a \to a^+_{\min}$), it does not approach $0$ but a finite positive value 
\begin{equation}
\label{eqn:smallWasym}
W_{\text{asym}}/N^2=\frac{4\varphi'^3}{9(12\pi^2-\varphi'^2)}~.
\end{equation}
Also, for the "big" black hole branch, we have an asymptotic relation 
\begin{equation}
    \label{eqn: bigWasym}
    W/N^2\sim -\frac{(\varphi'+2\sqrt{3}\pi)^3}{432}\tau^2.
\end{equation}
When $\varphi'=0$, it is in agreement with our benchmark case in subsection \ref{subsec:elemcase}.

\begin{figure}[htb]
    \centering
    \includegraphics[width=1\textwidth]{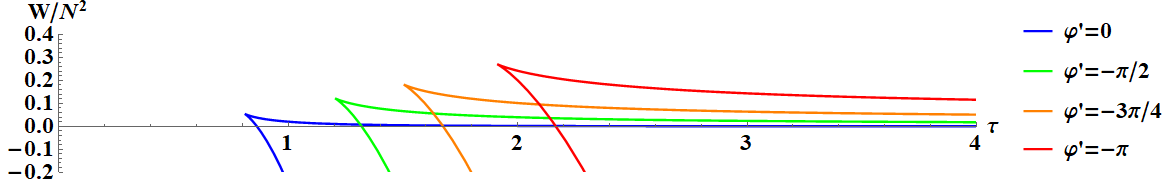}
    \caption{The BPS free energy $W$ as function of the BPS temperature $\tau$. The two angular momenta are equal ($\mu=0$) and $\varphi'=0,-\pi,-\frac{3\pi}{2},-2\pi$, from left to right. There are two branches in the phase diagram that meet in a cusp. Small black holes (the upper branch) asymptote to a positive BPS free energy 
    $W_{\text{asym}}$ \eqref{eqn:smallWasym} at large temperature when $\varphi'<0$.}
    \label{fig:W-tau}
\end{figure}

The phase diagram for various values of $\varphi'\leq0$ is presented 
in Figure \ref{fig:W-tau}.

The minimal temperature for a given $\varphi'$ is attained at $a_\text{cusp}$ determined by $\partial_a \tau = 0$ which leads to:   
\begin{equation}
    \varphi'=\frac{\pi (1-8a_\text{cusp} -11a_\text{cusp}^2)}{(2+a_\text{cusp})\sqrt{a_\text{cusp} (2+a_\text{cusp})}} ~.
    \label{eqn: cuspcond1}
\end{equation}
This equation is equivalent to a sextic in $a$ that cannot be solved analytically in general. For $\varphi'=0$, the condition \eqref{eqn: cuspcond1} reduces to a quadratic equation with the solution
$a_\text{cusp} = \frac{3\sqrt{3}-4} {11}$, in agreement with the result in subsection \ref{subsec:elemcase}. We also note that $a_{\text{cusp}}=1$ is a solution at the lower bound $\varphi' =\varphi'_-$ given in \eqref{eqn:varplim}. 
The BPS temperature $\tau$ and BPS free energy $W$ both diverge as $a\to 1^-$ so this shows that, as the lower bound is approached $\varphi'\to \varphi'_-$, the cusp moves to the far upper right corner in Figure \ref{fig:W-tau} and no physical black hole remain in the strict limit. 

Generally, the right hand side of \eqref{eqn: cuspcond1} is a monotonic function so, as $|\varphi'|$ increases from zero to its maximum $|\varphi'_-|$, $a_\text{cusp}$ increases monotonically through $[ \frac{3\sqrt{3}-4} {11} , 1[$. The BPS free energy \eqref{eqn:BPSGa=b} at the cusp moves as 
\begin{equation}
\label{eqn:Wcusp}
    \frac{dW_{\text{cusp}}}{da}=\left(\partial_aW\right)_{\varphi'}+\left(\partial_{\varphi'}W\right)_{a}\partial_a\varphi'=\left.\frac{\pi  a \left(3+14a-5 a^2\right)}{\sqrt{a (2+a)} (2+a)^2(1-a)^2}\right\vert_{a_{\text{cusp}}}>0~,
\end{equation}
with $\partial_a\varphi'$ evaluated from \eqref{eqn: cuspcond1}, so it increases monotonically as well. The BPS temperature \eqref{eqn:BPSTa=b} at the cusp increases monotonically entirely through its dependence on $\varphi'$ because it is a minimum $\partial_a\tau=0$. Taken together, these arguments establish analytically that the cusp moves monotonically up and to the right when $|\varphi'|$ increases. These trends are also visible in Figure \ref{fig:W-tau}.

We can analyze the Hawking-Page temperature similarly. The condition that the free energy $W=0$ yields the algebraic relation for $a_\text{HP}$:
\begin{equation}
 \varphi' = \frac{2\pi (1-5a_\text{HP}-2a^2 _\text{HP})}{(3-a_\text{HP})\sqrt{a_\text{HP}(2+a_\text{HP})}}~.
 \label{eqn: HPcond}
\end{equation}
The derivative of this expression
\begin{equation}
    \partial_{a_{\text{HP}}}\varphi'=-\frac{6 \pi  (1+a_{\text{HP}})^2 (1+3a_{\text{HP}})}{(3-a_{\text{HP}})^2 (a_{\text{HP}}(2+a_{\text{HP}}))^{3/2}}~,
    \label{eqn: aderivHPcond}
\end{equation}
is negative in the entire range $0\leq a\leq 1$. Therefore, the parameter $a_\text{HP}$ at the Hawking-Page transition depends monotonically on $\varphi'$. 
The corresponding temperature \eqref{eqn:BPSTa=b} moves as
\begin{equation}
\label{eqn:tauHPdel}
    \frac{d\tau_{\text{HP}}}{da}=
    \left(\partial_a\tau\right)_{\varphi'}+\left(\partial_{\varphi'}\tau\right)_{a}\partial_a\varphi' =\left. \frac{3 (1+3 a)}{\pi  (1-a)^3 \sqrt{a (2+a)}}\right\vert_{a_{\text{HP}}}> 0~, 
\end{equation}
so the Hawking-Page temperature $\tau_{\text{HP}}$ is a monotonically increasing function of $|\varphi'|$. This dependence is also apparent in Figure \ref{fig:W-tau}.

It is instructive to compare the phase diagram for $\varphi'$ in the range $(-2\pi\sqrt{3},0)$ with the benchmark case $\varphi'=0$ discussed in subsection \ref{subsec:elemcase}:
\begin{itemize}
    \item At a given BPS temperature $\tau$ the potential $|\varphi'|$ increases the BPS free energy $W$ on both branches. 
    \item The characteristic temperatures $\tau_{\text{HP}}$ and $\tau_{\text{cusp}}$ both increase with $|\varphi'|$. 
\end{itemize}
Both effects suggest an instability. Conversely, increased $\varphi'$ (decreased $|\varphi'|$) {\it stabilises} the BPS black hole. This is reminiscent of how an electric potential modifies the AdS-Reissner-Nordstr\"{o}m black hole, discussed in subsection \ref{subsubsec:phil3}.

\subsubsection{$\varphi'\geq0$}
\label{subsubsec:positivevarphi}

When $\varphi'>0$, the denominator in the second factor of $\tau$ \eqref{eqn:BPSTa=b} is strictly positive for all $a\in [0,1)$. Therefore, unlike when the potential $\varphi'<0$, the entire range $a\in (0,1)$ of the parameter $a$ is physical. 
At the lower end of the range $a=0$ the free energy $W=0$ and the temperature is 
\begin{equation}
\label{eqn:taumax}
\tau_{\rm max} = \frac{2}{\varphi'} ~.
\end{equation}
This is the maximal temperature on the small black hole branch, a novel feature of the $\varphi'>0$ regime. 
In the strict limit $a=0$ the geometry underlying the thermodynamic formulae reverts to pure AdS$_5$, it is not a black hole, so $\tau_{\rm max}$ is a bound, it cannot be reached. 
The bound on the temperature \eqref{eqn:taumax} is lowered when the potential $\varphi'$ increases. The range of allowed temperatures on the small black hole branch shrinks and, in the limit $\varphi'\to\infty$, it disappears altogether. In this limit the large black hole branch starts at $(W,\tau)=(0,0)$
and there is no small black hole branch. This limit is similar to the special case $\Phi=\Phi^*=3$ of ordinary (non-BPS) AdS-Reissner-Nordstr\"{o}m black holes that was discussed in subsection \ref{subsubsec:phi3}. 

\begin{figure}[H]
    \centering
    \includegraphics[width=1\textwidth]{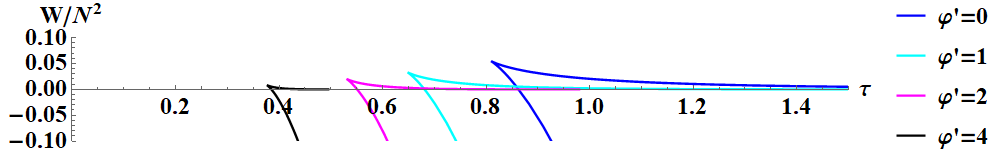}
    \caption{The free energy $\frac{W}{N^2}$ as function of BPS temperature $\tau$ for various $\varphi'\geq 0$. The values of $\varphi'= 0, 1, 2, 4$ increase from right to left. The small black holes (the upper branch) have $W=0$ at their maximal temperature $\tau_{\rm max}$ \eqref{eqn:taumax} which is $\tau_{\rm max}=\frac{1}{2}, 1$ for $\varphi'=4, 2$. The large black hole branch is qualitatively similar to its analogue for $\varphi'<0$ in Figure \ref{fig:W-tau}.}
    \label{fig:W-tau-1}
\end{figure}

The large black hole branch is not modified qualitatively from the benchmark case $\varphi'=0$ discussed in subsection \ref{subsec:elemcase}, or more generally the case $\varphi'\leq0$ developed earlier in this subsection. The phase diagram for various values of $\varphi'\geq0$ is plotted in Figure \ref{fig:W-tau-1}.

The figure indicates that, as we increase $\varphi'$ starting from zero, the coordinates
of the cusp $\tau_{\text{cusp}}$, $W_{\text{cusp}}$ both decrease, as does the Hawking-Page temperature $\tau_{\text{HP}}$ where the large black hole branch crosses $W=0$. To verify these features analytically, we note that the parameter $a_{\rm cusp}$, corresponding to the minimal temperature, is given by \eqref{eqn: cuspcond1} for either sign of $\varphi'$. It takes the value $a_\text{cusp} = \frac{3\sqrt{3}-4} {11}$ when $\varphi'=0$ and decreases monotonically when $\varphi'$ increases to positive values. The result follows because, for and $\varphi'$, the temperature $\tau$ \eqref{eqn:BPSTa=b} has positive derivative $\partial_{\varphi'}\tau>0$ and the motion  \eqref{eqn:Wcusp} of the free energy $W$. The estimate \eqref{eqn:tauHPdel} similarly establishes that $\tau_{\text{HP}}$ decreases as $\varphi'$ become larger. 

In summary, the evolution of the cusp, the Hawking-Page temperature, and the asymptotic behavior of the large black hole branch at high temperature, are all smooth for $\varphi'$ in the entire range $(\varphi'_-,\infty)$.
For these features the value $\varphi'=0$ plays no special role. In contrast, the asymptotic behavior on the small black hole branch depends sensitively on the sign of $\varphi'$: for $\varphi'\leq 0$, the temperature $\tau\to\infty$ is reached with an asymptotic value $W_{\rm asym}$ \eqref{eqn:smallWasym} of the free energy 
that is strictly positive except that it vanishes when $\varphi'=0$. For $\varphi'\geq 0$, the temperature is bounded by $\tau_{\rm max}$ \eqref{eqn:taumax} that is finite, except that it diverges when $\varphi'=0$.

\subsection{BPS Black Holes with Unequal Angular Momenta}
\label{subsec:BPS0mu}
In this paper we have presented nearly all formulae for general parameters $a, b$ but most examples focus on the ``equal angular momenta" case $a=b$. This is also the case that has been most studied in the literature, by far. In this subsection we elucidate the significance of ``unequal angular momenta" $a\neq b$ by taking the chemical potential $\mu \neq 0$. 

We will keep $\varphi'=0$ so the Gibbs energy $W$ \eqref{eqn:BPSGibbs} simplifies to
\begin{equation}
    \label{eqn:BPSGab}
    \frac{W}{N^2}=\frac{\pi (a+b)^2 \left[1-2a-2b-a^2-5ab-b^2-a^2b-ab^2\right] }{2(1-a)(1-b) \left(1+3a+3b+a^2+3ab+b^2\right)\sqrt{a+b+ab}}~,
\end{equation}
and the BPS potentials \eqref{eqn:BPST} and \eqref{eqn:BPSmu} read
\begin{align}
\label{eqn:BPSTab}
    &\tau=\frac{2 \left(1+3a+3b+a^2+3ab+b^2\right)}{  3\pi  (a+b)(1-ab)}\sqrt{a+b+ab}~,\\
\label{eqn:BPSmuab}
    &\mu= \pi \frac{a-b}{1+3a+3b+a^2+3ab+b^2}
    \frac{1+2a+2b+ab}{\sqrt{a+b+ab} }~.
\end{align}

\begin{figure}[H]
    \centering
    \includegraphics[width=0.9\textwidth]{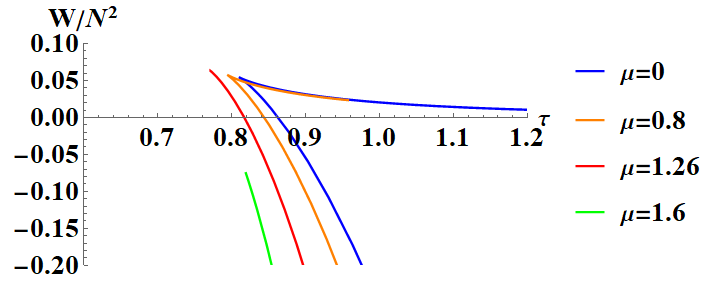}
    \caption{The BPS free energy $W$ vs. the BPS temperature $\tau$ for select values of the potential $\mu\geq 0$ that parametrizes the asymmetry between angular momenta. $\mu=0, 0.8, 1.26, 1.6$ from right to left. For small $\mu$, the phase diagram exhibits the familiar lower (large) and upper (small) black hole branches that meet in a cusp. For the small black hole branch, there is a maximal BPS temperature that decreases as function of $\mu$ and reaches the cusp at some $\mu=\mu_{\rm crit}$. The maximal value of $W$ also increases with $\mu\leq\mu_{\rm crit}$ but then decreases for $\mu>\mu_{\rm crit}$.}
    \label{fig:W-tau-mu0}
\end{figure}

Because of the antisymmetry of $\mu$ under the exchange $a\leftrightarrow b$, it is sufficient to analyze $\mu\geq0$. The phase diagram is shown in Figure \ref{fig:W-tau-mu0}. From the plot, for $\mu$ that is positive and small, the phase diagram evolves perturbatively from the benchmark case $\varphi'=\mu=0$ discussed in subsection \ref{subsec:elemcase} with some changes that appear smoothly:
\begin{itemize}
\item 
A maximal temperature of the small black hole branch develops that decreases when $\mu$ increases. The upper branch shrinks. 
\item 
For any given $\tau$, on either branch, the BPS potential $\mu>0$ lowers the free energy. This is expected because, when $a>b$ (\ref{eqn:BPSJa}-\ref{eqn:BPSJb}) give $J_a>J_b$, so the first law of BPS thermodynamics \eqref{eqn: firstlaw2} yields: $\partial_{\mu}W=-\frac{1}{2}(J_a-J_b)<0$.
\item The Hawking-Page transition temperature $\tau_{\text{HP}}$ also decreases as $\mu$ get larger, because both black hole branches are lowered. 
\item The maximal free energy $W_{\text{cusp}}$ increases with $\mu$, but the minimal BPS temperature $\tau_{\text{cusp}}$ decreases. Thus the cusp travels "towards North-West" as $\mu$ increases. 
\end{itemize}

Our interpretation of these features is that increasing $\mu$ is {\it destabilising}: it takes a lower temperature to achieve $W>0$ and so render the black holes unstable, and positive $\mu$  allows a higher maximal free energy $W_{\text{cusp}}$, indicating stronger instability. This is a BPS analogue of our finding in subsection \ref{subsubsec:phi0} that, for non-BPS black holes, angular velocities destabilise the AdS Reissner-Nordstr\"{o}m black holes.

The preceding comments only apply when the chemical potential $\mu$ is sufficiently small. From Figure \ref{fig:W-tau-mu0} we see that when $\mu$
exceeds a certain critical value $\mu_{\text{crit}}\approx1.26$ the small black hole branch disappears altogether, for such potentials only the ``large" black hole solution exists. The evolution when $\mu$ is larger than this critical value has the following features: 
\begin{itemize}
\item For any given $\tau$, increased $\mu$ lowers the BPS free energy. 
\item As $\mu$ increases, the minimal temperature $\tau_{\text{cusp}}$ increases as well but the maximal free energy $W_{\text{cusp}}$ decreases. 
\item 
The motion of the "cusp" shows that the values $\tau_{\text{min}}$ and $W_{\text{max}}$ evaluated at the cusp with $\mu=\mu_{\text{crit}}$ are the global minimum for the temperature and maximum for the BPS free energy, respectively. 
\item The Hawking-Page transition temperature $\tau_{\text{HP}}$ (where $W=0$) generally decreases as the entire large black hole branch is lowered. However, when the chemical potential is sufficiently large, above $\mu\approx1.5$, the entire large black hole branch is below $W=0$ and the transition disappears altogether. 
\item
As discussed in subsection \ref{subsec:genvarphimu}, the parameter $\mu$ satisfies an absolute upper bound $\mu_{\max}=\frac{3\pi}{5}\approx1.885$ when $\varphi'=0$. As $\mu\to\mu_{\max}$, \eqref{eqn:BPSmu} shows that $a\to1$ and $b=0$ so the BPS free energy \eqref{eqn:BPSGibbs} $W\to -\infty$. In the strict limit these is no underlying black hole solution. 
\end{itemize}

Here, we observe another effect of perturbing $\mu$ when $\mu$ is sufficiently large. Our interpretation is that for large $\mu$, it is {\it stabilising}: the entire black hole would lie below $W=0$ and thus render the black holes stable, and increasing $\mu$ lowers the maximal free energy $W_{\text{cusp}}$ indicating the stability. Since our discussion for non-BPS black holes is concentrated in the equal angular momenta case, we do not have an analogous case of it.

\subsection{Extreme Rotational Asymmetry: $\Omega_b'=0$}

As an extreme example of asymmetry between the two angular momenta, we consider the special case $\Omega_b'\equiv0$. In this case \eqref{eqn:Omegabphi} gives
\begin{equation}
    \label{eqn:varphipOmbp0}
    \varphi'=-\frac{2 \pi  \left(2a+b+a^2+2ab\right)}{(1+b) \sqrt{a+b+ab}}~.   
\end{equation}
It follows that that $\varphi'$ must be negative. Formally, we can also achieve $\Omega_b'=0$ by taking $b=1$, but then the free energy $W$ \eqref{eqn:BPSGibbs} diverges, so we dismiss this possibility as an unphysical limit. 

\begin{figure}[H]
    \centering
    \includegraphics[width=0.9\textwidth]{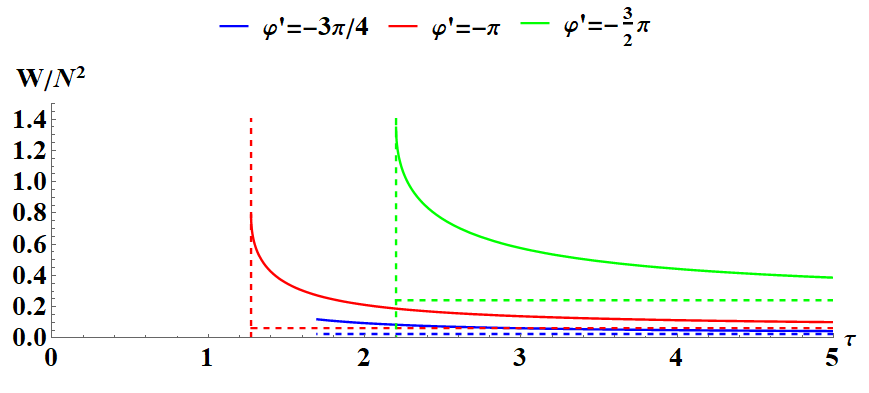}
    \caption{$W$ vs. $\tau$ when $\Omega_b'=0$ with $\varphi'=-\frac{3}{2}\pi$, $-\pi$ and $-\frac{3}{4}\pi$ (upper to lower respectively). There are only "small" black holes:  they all have $W>0$ and negative specific heat. For any given $\varphi'$, there is a minimal BPS temperature which corresponds to the maximal BPS free energy.}
    \label{fig:W-tau-mu-Ob=0}
\end{figure}

When $\Omega_b' = 0$ the BPS temperature \eqref{eqn:BPST} and the BPS free energy \eqref{eqn:BPSGibbs} simplify
\begin{equation}
\label{eqn:tauWOmbp0}
    \begin{split}
        \tau&=-\frac{2}{\Omega_a'}=-\frac{2}{\mu}=\frac{2 (1+b) \sqrt{a +b+ab}}{\pi  (1-a) (b-a)}~,\\ 
        W&=\frac{\pi  (1+a) (a+b)^2}{2 (1-a) (1+b) \sqrt{a+b+ab}}~.
    \end{split}
\end{equation}
We see that $W$ is positive definite. The phase diagram of $\tau$ and $W$ for various fixed $\varphi'$ is presented in Figure \ref{fig:W-tau-mu-Ob=0}. There are several features:
\begin{itemize}
    \item For any given $-2\sqrt{3}\pi<\varphi'<0$, 
    the BPS free energy $W$  decreases monotonically with $\tau$.  
    \item The lower bound $\tau_{\min}$ decreases in the range $-2\sqrt{3}\pi<\varphi'<-\pi$ and increases when $-\pi<\varphi'<0$. It  reaches its global minimum $\tau_{\min}=\frac{4}{\pi}$ at $\varphi'=-\pi$. 
    \item $W$ displays vertical asymptotes at $\tau_{\min}$ as long as $b$ can approach each $1$ (allowed for $-2\pi \sqrt{3}<\varphi'<-\pi$), whereas for $\varphi'>-\pi$, $W$ reaches an upper bound but with no asymptote (such as the $\varphi'=-\frac{3}{4}\pi$ curve in Figure \ref{fig:W-tau-mu-Ob=0}).
    \item $\tau$ diverges when $a$ approaches $b$ from below, leading to an asymptotic value for $W$ at large $\tau$: $\frac{2\varphi'^2}{9(\varphi'^{2} -12\pi^2)}$. 
\end{itemize}

In summary, the large black hole branch disappears entirely when $\Omega_b'=0$, there are only small black holes. This is reminiscent of the general (not BPS) black holes with $\Omega=1$, discussed in subsection \ref{subsubsec:Omega=1}. One aspect of the special case $\Omega_b'=0$ is that it corresponds to ``maximal asymmetry", there is a BPS potential for rotation along the ``a" direction, but none along ``b". However, a more illuminating perspective is that the definition of the ``primed" potentials \eqref{eqn: primedlmt} shows that when the BPS potential $\Omega_b'=0$, the physical rotational velocity $\Omega_b=1$ reaches the speed of light. From this point of view the close analogy with the $\Omega=1$ non-BPS black holes is expected.



\section{Discussion}
In this article, we studied the thermodynamics of AdS black holes via analysing their phase diagrams. Compared with previous studies \cite{Chamblin:1999tk, Caldarelli:1999xj}, this work has a particular emphasis on the role of rotation. We pointed out that rotation tends to {\it destabilise} a black hole in the sense that rotation increases the maximal free energy. 

We also developed BPS thermodynamics systematically and, in many explicit examples, we pointed out that the phase diagram of the BPS black hole exhibits the similar "cusp" structure as the well-known AdS Schwarzschild black hole phase diagram. We emphasised the role of an important fugacity, $\varphi'$, that preserves BPS saturation. This fugacity $\varphi'$ has been set to zero in the BPS limit in previous discussion \cite{Choi:2019miv,Kinney:2005ej}, and paid little attention. We illustrated how $\varphi'$ brings qualitative change of the phase diagram. Besides, we studied the case where the BPS black holes carry two unequal angular momenta, and discovered a qualitative change of the phase diagram. 

However, there are many open questions. For example, we introduced a notion of temperature for BPS black holes in \eqref{eqn:BPST} as a generalisation to the work by {\it Choi et al.} in \cite{Choi:2019miv}. It is for sure that one should be able to obtain a deeper understanding on this BPS temperature. Another important question is the physical meaning of $W=0$. Is this phase transition true or, instead, an illusion. 
We look forward to pursue these and related questions in future research work.

\section*{Acknowledgements}

This research was supported in part by the U.S. Department of Energy under grant DE-SC0007859. NE is supported in part by a Leinweber Graduate Fellowship.

\bibliographystyle{JHEP}
\bibliography{main.bib}


\end{document}